\documentclass[sigconf]{acmart}
\usepackage{graphicx}
\usepackage[lofdepth,lotdepth]{subfig}
\usepackage{csquotes}
\usepackage{url}
\usepackage{adjustbox}
\AtBeginDocument{%
  \providecommand\BibTeX{{%
    \normalfont B\kern-0.5em{\scshape i\kern-0.25em b}\kern-0.8em\TeX}}}

\copyrightyear{2021}
\acmYear{2021}
\setcopyright{acmcopyright}\acmConference[CIKM '21]{Proceedings of the 30th ACM International Conference on Information and Knowledge Management}{November 1--5, 2021}{Virtual Event, QLD, Australia}
\acmBooktitle{Proceedings of the 30th ACM International Conference on Information and Knowledge Management (CIKM '21), November 1--5, 2021, Virtual Event, QLD, Australia}
\acmPrice{15.00}
\acmDOI{10.1145/3459637.3482395}
\acmISBN{978-1-4503-8446-9/21/11}

\begin{document}
\title{Speaker-Conditioned Hierarchical Modeling for Automated Speech Scoring}
\author[]{Yaman Kumar Singla\textsuperscript{1,2,3*}, Avyakt Gupta\textsuperscript{1*}, Shaurya Bagga\textsuperscript{1}}
\authornote{Equal Contribution}
\author[]{Changyou Chen\textsuperscript{3}, Balaji Krishnamurthy\textsuperscript{2}, Rajiv Ratn Shah\textsuperscript{1}}
\affiliation{
  \institution{1. IIIT-Delhi, 2. Adobe, 3. SUNY-Buffalo}
  \country{1. India, 2. India, 3. USA}
}

\renewcommand{\shortauthors}{Singla et al}

\begin{abstract}
 Automatic Speech Scoring (ASS) is the computer-assisted evaluation of a candidate's speaking proficiency in a language. ASS systems face many challenges like open grammar, variable pronunciations, and unstructured or semi-structured content. Recent deep learning approaches have shown some promise in this domain. However, most of these approaches focus on extracting features from a single audio, making them suffer from the lack of speaker-specific context required to model such a complex task. We propose a novel deep learning technique for non-native ASS, called speaker-conditioned hierarchical modeling. In our technique, we take advantage of the fact that oral proficiency tests rate multiple responses for a candidate. We extract context vectors from these responses and feed them as additional speaker-specific context to our network to score a particular response. We compare our technique with strong baselines and find that such modeling improves the model's average performance by 6.92\% (maximum = 12.86\%, minimum = 4.51\%). We further show both quantitative and qualitative insights into the importance of this additional context in solving the problem of ASS.

\end{abstract}




\maketitle
\definecolor{mygreen}{HTML}{0ED900}

\newcommand{\shaurya}[1]{\textcolor{blue}{[Shaurya]: #1}}
\newcommand{\yaman}[1]{\textcolor{red}{[Yaman]: #1}}
\newcommand{\avyakt}[1]{\textcolor{mygreen}{[Avyakt]: #1}}

\section{Introduction}
\label{sec:Introduction}
Automated Scoring (AS) of language is one of the most popular applications of artificial intelligence. The first such reporting goes back to 1960s for the Project Essay Grade (PEG) by Ellis Page \cite{page1967statistical}. Since then, the systems have expanded by leaps and bounds and now they grade millions of test-takers per year \cite{laflair2019duolingo,educational2014snapshot,OhioNumbers}. The scores given by them are used in important decisions including college admissions, visa approvals, screening interviews, and even high-school exams \cite{ohioAES2,OhioNumbers,UtahNumbers,educational2014snapshot}. Apart from its immense importance in the education and learning domain, automatic scoring also has a large market size of more than USD 110 billion, with a US market size alone of USD 17.1 billion \cite{researchAndMarketsEdTesting,causeIQETSWorth,educationTestingSalaries,ibisWorldEdTesting}.

AS, in general, refers to the act of using computers to convert a candidate's performance on standardized tests to some performance metrics. AS systems are required to interpret and analyse the candidate's response and make a prediction that can be used to infer the person's ability \cite{ESKENAZI2009832}. Automatic Speech Scoring (ASS), a type of AS, assesses the speaking proficiency of a candidate in a language. ASS systems are used for a variety of reasons such as to decide admissions to universities, alleviate the workload of teachers, save time and costs associated with grading, and evaluation of online courses. For instance, a British teacher spends an average of 8 hours per calendar week scoring exams and assignments \cite{higton2017teacher}. This figure is even higher for developing and low-resource countries where the teacher to student ratio is dismal \cite{oecd2014indicator}. ASS systems can effectively reduce this workload while still maintaining quality \cite{kumar2019get}. Due to this reason, schools in Ohio and Utah have already started automated essay scoring for their high school students and would soon require speech scoring systems \cite{ohioAES2,OhioNumbers,UtahNumbers}. Additionally, with the rise of people taking online courses, online platforms require ASS systems to provide immediate results and feedback to the user. On account of all these reasons, much research and development is required in ASS to better the systems which play such an important part in life-changing decisions including university admissions, visas, and job interviews.

Currently, the majority of work in deep learning based AS has been done in the Automatic Essay Scoring (AES) domain. 
Speech scoring is a much harder problem than essay scoring as we have to model both the content (\textit{what} was spoken) and delivery (\textit{how} it was spoken). We also have to deal with open grammar, unstructured or semi structured content, prosody, fluency and pronunciation variations. Moreover, it has been noted in previous studies that applying AES approaches for ASS results in significant performance drops \cite{loukina2016automated} due to the mentioned issues. To deal with these challenges, most of the previous studies have used handcrafted features for scoring speech. For instance, TOEFL\textsuperscript{\textregistered} from Education Testing Service (ETS) is a well-known example which uses a rich set of handcrafted features for ASS \cite{speechrater}. However, handcrafted rules are unable to model higher level features such as opinion formation, arguments, prose coherence, \textit{etc}. Therefore, recently more and more approaches are moving towards end-to-end deep learning models for scoring speech responses. For instance, \citet{chen2018end} combine acoustic and lexical models to perform ASS. \citet{qian2018prompt} and \citet{grover2020multi} use LSTM-attention and attention fusion, respectively, to score spontaneous speech responses. Recently, research studies have also shown that incorporating speech transformers for scoring improves performance \cite{wang2021automated,grover2020multi}.

Contrary to the true classroom settings, a majority of existing feature engineering and deep-learning research treats automatic speech scoring of samples of the same candidate on different prompts\footnote{Here a prompt denotes an instance of a unique question asked to test-takers for eliciting their opinions and answers in an exam. The prompts can come from varied domains including literature, science, logic and society. The responses to a prompt indicate the creative, literary, argumentative, narrative, and scientific aptitude of candidates and are judged on a pre-determined score scale.} to be independent of each other \cite{bejar2017threats,yan2020handbook}. That is why most of the previous literature reports the performance of \textit{one model per prompt} with no or minimal information sharing between two models. It is well-known that a speaker's performance on different prompts is related. Even in classroom settings, while scoring a response, a teacher knows the subtleties of each pupil on account of her knowledge of the pupil's performance on previous occasions \cite{bejar2017threats,yan2020handbook}. Therefore, conditioning models on speakers can lead to information sharing across models corresponding to different prompts and hence is expected to improve the models' performance.

\begin{figure}[h]
    \centering
    \includegraphics[width=6.5cm]{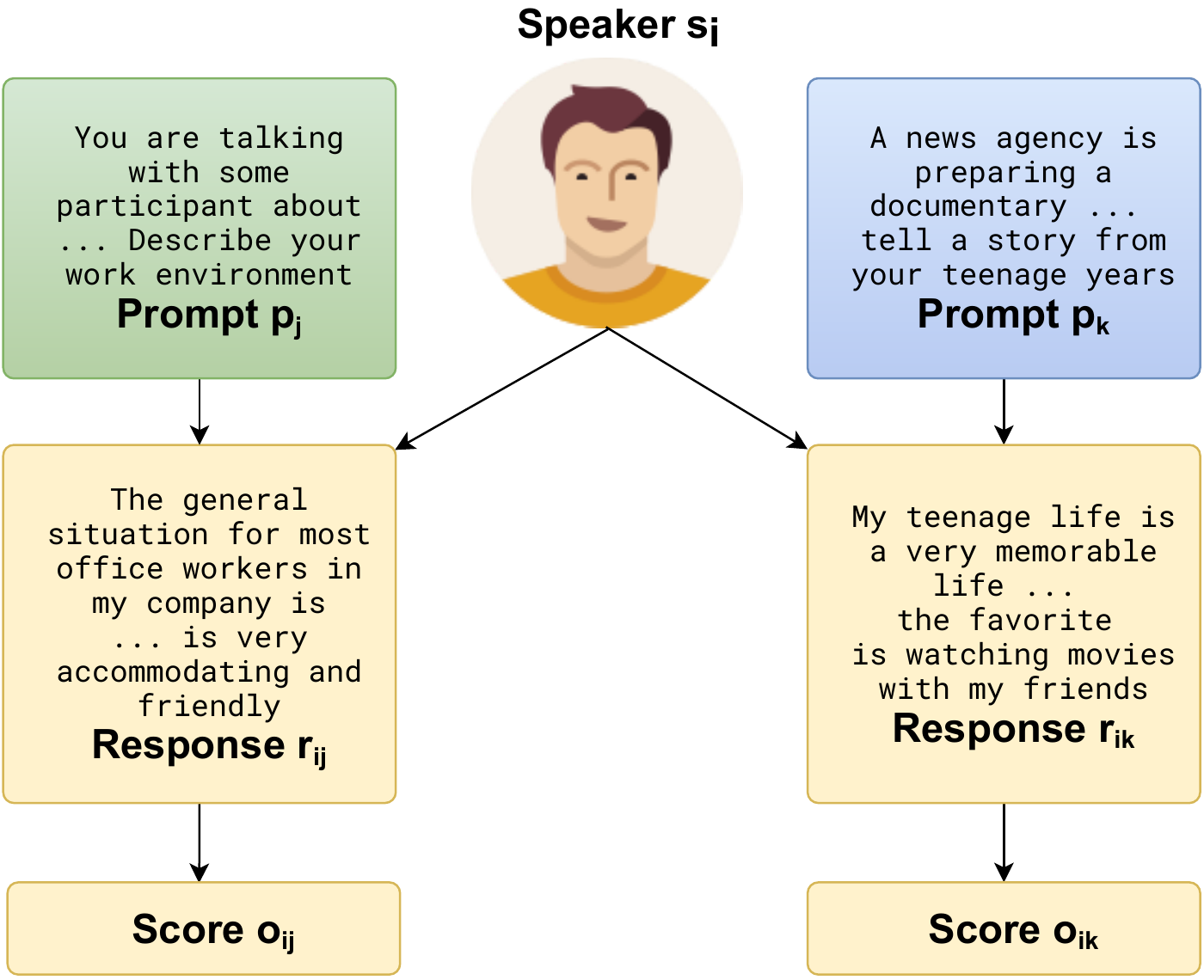}
    \caption{\small Speaker-Prompt Interaction: Motivation for speaker-conditioned models. Speaker $s_i$ gives responses $r_{ij}$ and $r_{ik}$ on prompts $p_j$ and $p_k$ and obtain the score $o_{ij}$ and $o_{ik}$.
    }
    \label{fig:speaker-conditioned models}
\end{figure}

Fig.~\ref{fig:speaker-conditioned models} presents the basic idea of speaker-prompt interaction. Response $r_{ij}$ is determined by two factors: speaker $s_i$ and prompt $p_j$. Speaker $s_i$ remains constant while prompt $p_j$ changes for generating (answering) a response $r_{ij}$. Therefore, a model finding the score $o_{ij}$ for a response $r_{ij}$ can benefit from sharing speaker $s_i$'s information with other prompts ($p_k$ s.t. $k\neq j$).
The same idea of information sharing through speaker conditioning has been explored directly or indirectly in other tasks including speaker activity detection \cite{ding2019personal}, speaker diarization \cite{rouvier2015speaker}, text-to-speech \cite{hsu2018hierarchical}, speech translation \cite{mccarthy2020skinaugment}, and speaker detection \cite{klusacek2003conditional} in the speech domain, and authorship verification \cite{king2020authorship}, author-based predictive writing \cite{ji2019learning}, sentiment analysis \cite{yang2017overcoming}, and natural text generation \cite{oraby2018controlling} in the NLP domain. Conditioning on speakers improved the performance on all these tasks.

Building on these gaps in existing research, we propose a novel deep learning architecture which relies on speaker-conditioned hierarchical modeling, where the prompts of each user are bundled by the user-specific context. Specifically, we feed our per-prompt models with user-specific context,  allowing the models to better understand the user and make a more informative prediction (\S\ref{sec:Modeling strategies}). Using an exhaustive set of experiments involving 108 unique model-prompt pairs, we compare our technique with strong baselines and show the superiority of the proposed speaker-conditioned hierarchical models (\S\ref{sec:results}). We also show that audio features along with text-based features can boost the model performance, as audio features capture details such as stress, hesitation, word intensity, word duration, \textit{etc.} which are not directly captured by text transcripts and hence not modeled by the text based models. 
In \S\ref{sec:quantitative analysis} we analyze quadratic weighted kappa and mean squared error to see how well our models perform in comparison to the baselines. We analyze how conditioning on other prompts changes the predictions for high-bias examples. We also note the speaker level accuracy increase by conditioning on speaker for the proposed models. Later, in \S\ref{sec:Qualitative Analysis}, we interpret the results to see how the speaker-conditioned hierarchical models obtain better predictions than the baseline models. For this, we analyze the information-sharing strengths across prompts (\S\ref{sec:Qualitative Analysis})

The main contributions of our work are summarised as follows:
\begin{itemize}
    \item We propose two types of speaker-conditioned hierarchical models for the problem of ASS. To the best of our knowledge, this is the first time speaker conditioning is considered on a multi-prompt dataset in ASS.

    \item Through extensive experiments on 108 unique model-prompt pairs on a real-life non-native speaker dataset, which contains 7,198 speakers answering 6 prompts each, we show that the proposed models achieve state-of-the-art results on speech scoring. Our method increases average human-machine agreement by 6.92\% (maximum = 12.86\%, minimum = 4.51\%) over strong baselines. 
    
    \item Further, using interpretability experiments, we show how our model learns the relations between prompts and individual speaker responses via speaker-conditioned hierarchical embeddings. We find that this also increases the precision for all prompt-specific models. We also find the relative importance of audio and text in ASS.
\end{itemize}


\begin{table*}[htbp]
 \begin{tabular}{|c|c|c|c|c|c|} 
 \hline
 \textbf{Prompt}  & \textbf{Difficulty} & \textbf{Avg \#sec.} & \textbf{Avg \#tokens} & \textbf{Score Distribution} \\
 \hline
 1 & B1 & 57.79 & 101.91 & A2(197)/Low B1(1286)/High B1(5715) \\ 
 \hline
 2 & B1 & 58.82 & 111.18 & A2(400)/Low B1(2697)/High B1(4101) \\
 \hline
 3 & B2 & 81.86 & 151.58 & A2(65)/Low B1(463)/High B1(3087)/Low B2(3497)/High B2(86) \\
 \hline
 4 & C1 & 104.83 & 183.83 & A2(65)/LB1(463)/High B1(3087)/LB2(3497)/High B2(86) \\
 \hline
 5 & C1 & 106.67 & 199.61 & A2(60)/LB1(382)/High B1(2675)/LB2(3945)/High B2(136) \\ 
 \hline
 6 & B1 & 56.15 & 111.22 & A2(56)/Low B1(767)/High B1(6375) \\
 \hline
\end{tabular}
\caption{\label{tab:dataset} \small Statistics of the SOPI-SLTI dataset. Difficulty measures the CEFR scale difficulty of each prompt. We report the distribution of each prompt across proficiency levels in the "Score Distribution" column.}
\end{table*}


\section{Related Work}
\label{sec:related work}

\textbf{Automatic Speech Scoring (ASS):} In general, ASS can be of the following types: read-aloud, repeat-aloud, keyword sentence completion, structured narrative, providing an opinion, listen-speak, and conversation-based assessments \cite{evanini2017approaches}. Amongst them, structured narrative, providing an opinion, and conversation-based assessments are the most thorough and hence the most challenging ones \cite{evanini2017approaches}. In this work, we cover the first two of these three. The range of features covered by this type of assessment is highly comprehensive and include fluency, grammar, vocabulary, discourse coherence, and content, amongst others \cite{xiong2013automated}. Due to the complexity of the task, earlier systems relied on simply measuring automatic speech recognition (ASR) errors on already fixed set of sentences \cite{WITT200095}.

Over the years, a lot of studies have focused on extracting handcrafted features that cover different aspects of speech such as fluency, rhythm, intonation and stress, vocabulary use, \textit{etc.} \cite{speechrater}. However, these systems are not able to capture complex high-level features such as opinion formation, structure of prose, argument depth, \textit{etc}. Studies have shown that end-to-end systems work better than feature-based systems, while also obviating the need for feature engineering. 

\citet{Chen2018EndtoEndNN} were the first to propose an end-to-end network that captures both lexical and acoustic cues to score samples. They used Bi-directional LSTM with attention to generate features from text, and for acoustic cues they used Kaldi ASR's \cite{Povey_ASRU2011Kaldi} outputs to obtain acoustic model posterior probabilities and word durations, and Praat \cite{Boersma2009praat} to obtain pitch and intensity values. They reported an improvement over the conventional method of handcrafted features \cite{speechrater}. \citet{saeki2021analysis} proposed a combination of three models: lexical, acoustic and visual models to score conversation based assessments. \citet{grover2020multi} proposed a multimodal end-to-end system, which uses a Bi-directional Recurrent CNN and Bi-directional LSTM to encode acoustic and lexical cues from spectrograms and transcriptions, respectively. They applied attention fusion on these features to learn complex interactions between different modalities before final scoring and reported consistent empirical improvements of their system over strong baselines. \citet{qian2019neural} showed improvements over baselines after including prompt-level information in their models.

\textbf{Automated Essay Scoring (AES):} A task related to ASS is AES. It is one of the most important educational applications of natural language processing (NLP) in the education domain \cite{thomas2016future,ohioAES2,UtahNumbers}.
The first neural network approach was proposed by T$\&$N \cite{taghipour-ng-2016-neural}. They proposed a network which uses Convolutional Neural Network (CNN) to encode local context information, and a long short term memory network (LSTM) to encode long term dependencies. Dong \cite{dong-etal-2017-attention} improved this by including the attention mechanism into the network by using attention pooling of CNN features instead of the simple max pooling/average pooling. \citet{tay} hypothesized that scoring essays can be improved by computing and using coherence scores of an essay, since coherence is an important dimension of essay quality. They modeled coherence by adding an additional layer to their network which takes in two hidden states of the LSTM network and outputs the similarity or coherence. They used these coherence values to enrich their features and boost their performance.

However, none of the approaches in both AES and ASS have looked into the possible gains after conditioning them on speakers. Our work differs from the current works as we explore speaker-conditioned hierarchical modeling, a way to enrich our feature space by providing speaker-specific cues obtained from previously trained models. It has been shown by various previous studies that speaker-conditioned models get better performance across many tasks such as text-to-speech and natural text generation \cite{hsu2018hierarchical,oraby2018controlling}.


\begin{figure*}[h]
    \centering
    \subfloat[][One Stage speaker-conditioned hierarchical model - Here we use the context vectors $c_{ik}$ from $E_{k}$ of all the prompts $p_{k}$ where $k < j$ along with $c_{ij}$ to predict the output score $o_{ij}$.]{
    \includegraphics[width=0.9\columnwidth]{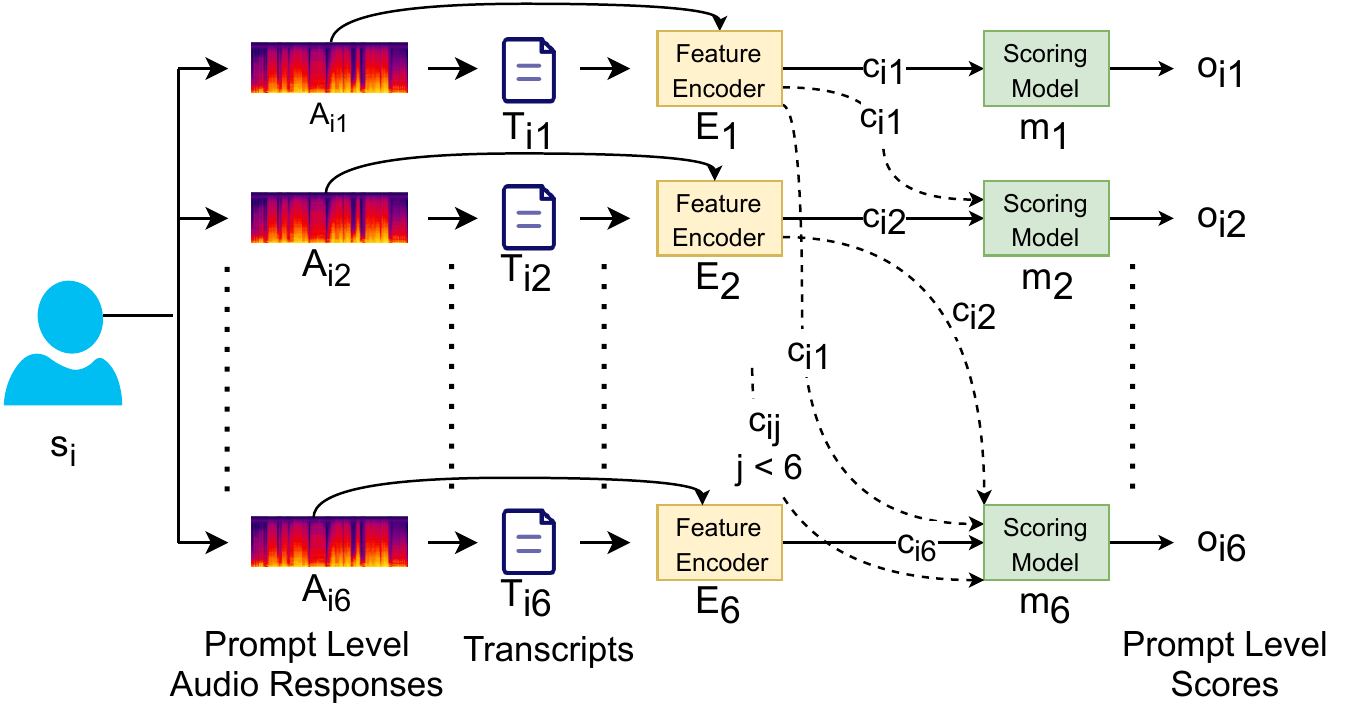}
    \label{fig:hierarchical}}
    \qquad
    \subfloat[][Two Stage speaker-conditioned hierarchical model  - Here we use the context vectors $b_{ik}$ from baseline encoder $E_{k}$ of all the prompts $p_{k}$ where $k \neq j$ along with $c_{ij}$ to predict the output score $o_{ij}$.]{
    \includegraphics[width=0.9\columnwidth]{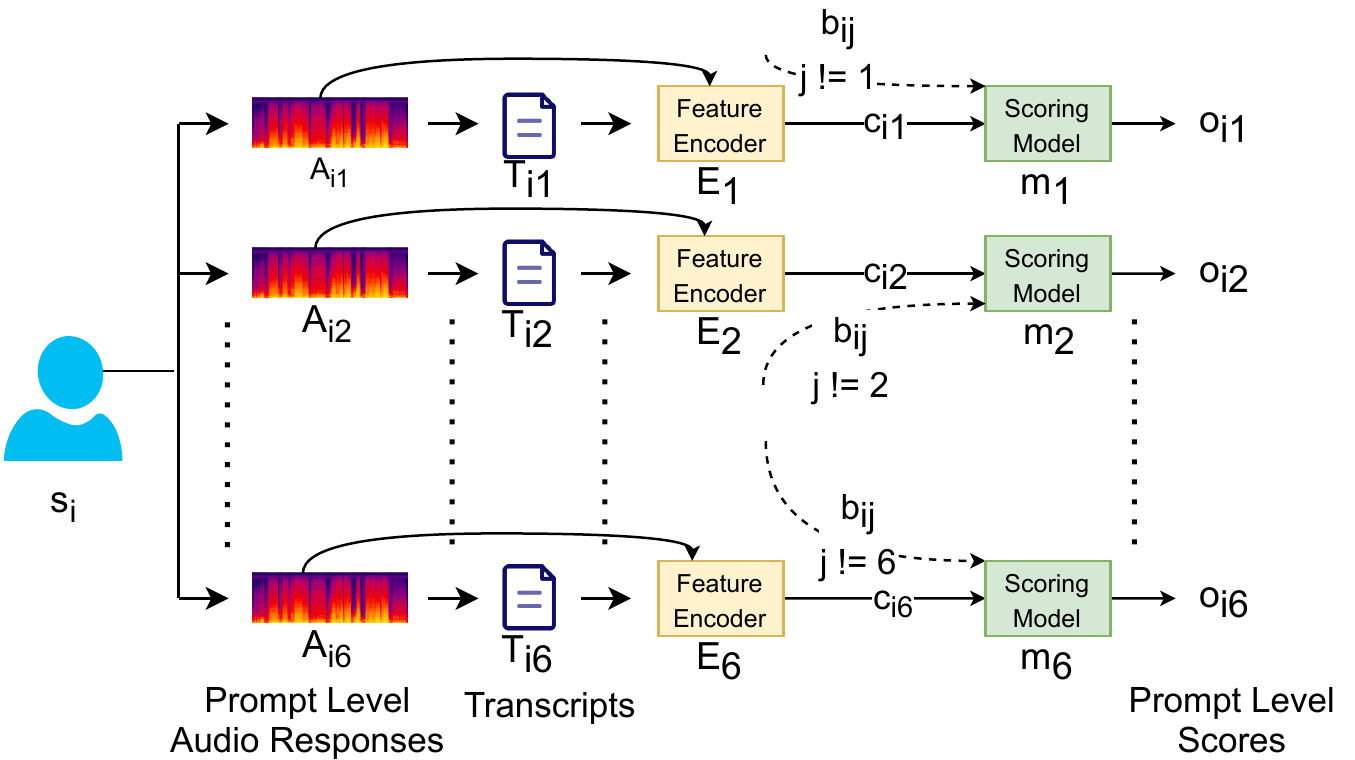}
    \label{fig:two stage hierarchical model}}
    \caption{Speaker-conditioned hierarchical models. Above figures show scoring models ($m_j$) scoring audio responses $a_{ij}$ for speaker $s_i$ on prompt $p_j$.}
    \label{fig:hierarchical models}
\end{figure*}

\begin{figure}[htbp]
    \centering
    \includegraphics[width=8cm]{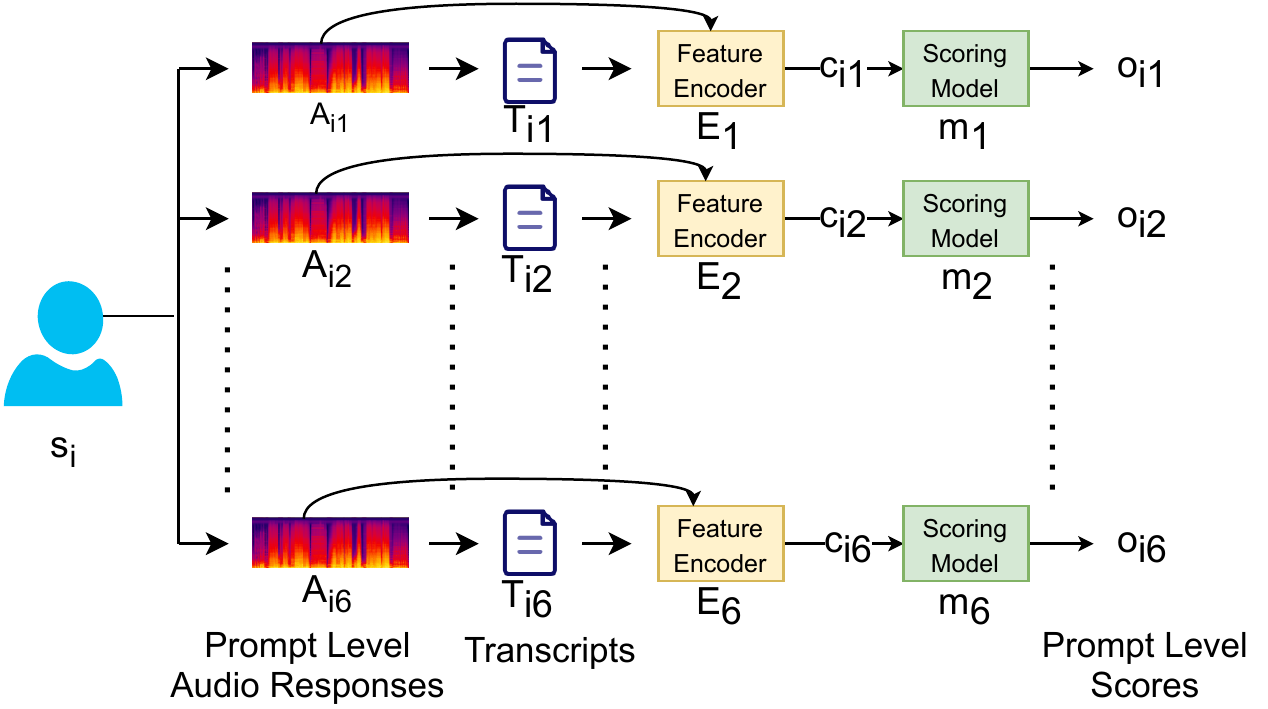}
    \caption{\small Baseline Modeling Strategy. The figure shows scoring models ($m_j$) scoring the audio responses $a_{ij}$ for the speaker $s_i$ on prompt $p_j$. The audio responses $a_{ij}$ are converted to deep embeddings $c_{ij}$ using text and audio encoders $E_j$.}
    \label{fig:baseline}
\end{figure}

\section{Task and Dataset}
\label{sec:Task and Dataset}

\subsection{Dataset}
\label{sec:Dataset}
In this study, we use the data collected by Second Language Testing Inc. (SLTI) while administering the Simulated Oral Proficiency Interview (SOPI) Exam for L2 English speakers \cite{grover2020multi,stansfield1992development}. The SOPI exam has been operational since 1992 and has a rich research history  \cite{stansfield1992development,stansfield1996test}. Currently, SOPI is used for screening interviews, university admissions, employee training and skill development, university and job placement, and as a test in several online courses \cite{sopi-slti}. The SOPI offers psychometric advantages in terms of reliability and validity, particularly in standardized testing situations \cite{malone2000simulated}. 

A majority of the speakers in the released dataset are from the Philippines region. The candidates are high school graduates and above. A SOPI test-taker is presented with a question paper containing six prompts on their computer screens, and their responses for each individual item are recorded. The test taker is given approximately one minute to think and two minutes to respond to each question. To answer the questions in the form, the speaker needs to provide arguments and explanations supporting their opinion.

These responses are rated by two expert raters, and in case of a disagreement on the overall score a third rater is asked to resolve the same. The prompts and the rubrics for evaluation follow the guidelines as proposed by the Central European Framework of Reference for Languages (CEFR) \cite{cefr}. CEFR is an international standard for describing language ability. It describes the language proficiency on a six-point scale, from A1 for beginners to C2 for those who are proficient in the language.

The levels of questions vary from B1 to C1, and the ratings provided to the speakers are upper bounded by the level of the question. The distribution of the dataset and the relevant statistics is provided in Table~\ref{tab:dataset}. The dataset consists of 931 hours of speech spoken by 7,198 speakers. The test elicits argumentative, narrative, persuasive and expository skills to answer the questions. For example, the prompt 1 asks candidates to describe their work environment. A candidate gets two types of scores: prompt-level and overall (global) score. Overall score is computed by combining all the prompt-level scores. Other studies also use the data for various purposes including automated scoring, coherence, \textit{etc.} \cite{grover2020multi,patil2020towards,bamdevAutomated,smith2017testing,stansfield2008testing,stansfield1992research}.


\subsection{Experimental Setup}
\label{sec:Experimental Setup}
We stratify our dataset based on the speaker and global score and then split the candidates in train, validation and test set in 70:10:20 ratio. This ensures that for each prompt we have no speaker intersection in the train, validation and test sets, as required by our speaker-conditioned hierarchical modeling approach. It is noteworthy that this also reflects the real deployment setting. For each prompt, we train our model and report the quadratic weighted kappa (QWK) score on the test set.

We use DeepSpeech2 \cite{deepspeech2} based ASR system with a trigram language model for transcription of the non-native responses. The ASR system is trained on approximately 1000 hours of audio sampled from CommonVoice \cite{commonvoice} and LibriSpeech datasets \cite{librispeech} and further fine-tuned on approximately 22 hours of transcribed non-native spoken responses from our dataset. The dataset is further augmented with noise using AudioSet \cite{gemmeke2017audio}. This ASR achieves a word error rate of 16.63\%.

We tokenize the transcripts using the Spacy tokenizer \cite{spacy2} and lowercase them. Words not part of the training data vocabulary are mapped to the unknown token (UNK) and initialized to zero vector in the embedding layer. We treat the scoring of the responses as a regression problem such that the output $o_{ij}$ is the normalized score of the speaker $s_i$'s response $r_{ij}$ to a prompt $p_j$. The CEFR \cite{cefr} proficiency levels (N levels) associated with the responses are normalized to the range [0, 1] for training. While testing, we rescale the model output back to the original score range and measure the performance. All our experiments are done using PyTorch \cite{pytorch}. All the plots and hyperparameter tuning were assisted by Weights and Biases \cite{wandb}. Next, we present our models that were used as feature extractors and the proposed modeling strategies.

\subsection{Models}
\label{sec:models}
We experimented with several model architectures including Bi-directional LSTM (BDLSTM), Bi-directional LSTM with Attention (BDLSTM+Attn), and BERT based model. We also tried including audio features using wav2vec2.0 \cite{wav2vec2} with each text encoder model. We demonstrate each of these models with 3 modeling strategies thus showing results for 18 (6*3) different models for 6 prompts. In total, we experiment with 108 (18*6) individual model-prompt pairs. The three modeling strategies are: baseline strategy without any speaker-conditioning (Fig.~\ref{fig:baseline}), one stage speaker-conditioned hierarchical modeling (Fig.~\ref{fig:hierarchical}), two stage speaker-conditioned hierarchical modeling (Fig.~\ref{fig:two stage hierarchical model}). Before explaining the different modeling strategies in \S\ref{sec:Modeling strategies}, we first give a brief overview of each of the models.

\textbf{Bi-Directional LSTM:}
The Bi-directional LSTM is capable of learning long-term dependencies. A bi-directional network is chosen here to take into account information both from the past and future given the inherent nature of speech and language production. Our model configuration is similar to \cite{Chen2018EndtoEndNN}. The words from the preprocessed transcripts are mapped to their embedding. This word embedding layer is initialized with 300-Dimensional glove embeddings \cite{glove} trained on Wikipedia and is optimized during training. We use a single layer of BDLSTM, and assign the hidden layer from the last time step as the context vector ($c$) for the entire response.

\textbf{BDLSTM+Attention:} The attention mechanism \cite{attention} has achieved state-of-the-art on various natural language processing tasks. It enables weighting contextual information learned during each time step, allowing the model to determine which states to pay attention to. We apply attention to the hidden states of a BDLSTM. Given the hidden states of the BDLSTM \textbf{$h$}, we calculate the context vector \textbf{$c$} as follows:

\begin{equation}
\begin{aligned}
e_t = W_ah_t\\
a_t = softmax(e_t)\\
c = \sum_{i=1}^{T}h_ta_t\\
\end{aligned}
\end{equation}

where $h_t$ is the hidden representation at the $t^{th}$ timestep, $W_a$ is the weight matrix for the attention layer which assigns importance scores ($e_t$) to each hidden state. These scores are then normalised to sum to 1 ($a_t$) using a softmax layer. The context vector $c$ is a convex combination of the hidden states with weights $a_t$.

\textbf{BERT:} We use the BERT Tokenizer and the BERT Base model from the huggingface library \cite{huggingface}. Pretrained BERT \cite{bert} based embeddings have set a new standard in many natural language processing tasks.  The BERT base model has 12 multiheaded self-attention layers with a total of 110M trainable parameters. It produces a 768-dimensional feature vector for a text sequence. We fine tune the last 7 self-attention layers and add a projection layer on top of the features extracted to obtain the context vector ($c$) for a response.

\textbf{Multi Modal Models:}
We hypothesise that audio features, along with text-based features can improve the performance of models. We rely on the previous studies that show that audios capture features like stress, hesitation, rhythm, word duration, pitch, intensity, \textit{etc.} which are absent in transcripts and hence text-based features do not give complete information required for scoring \cite{shah2021all}. To extract features from audios, we use the pretrained wav2vec2.0 \cite{wav2vec2} model provided by the huggingface library \cite{huggingface}. Features extracted from wav2vec2.0 have shown SOTA results on various speech tasks \cite{shah2021all}.

Wav2vec2.0 takes in the raw audio sampled at 16khz, and outputs a 768 dimensional feature vector ($v^{a}$), summarising the entire audio.
We store these features ($v^{a}$) of every audio extracted from wav2vec2.0 and apply a fully connected layer over these features to finetune them and obtain the audio context vector $c^{a}$.
\begin{equation}
    c^{a} = W^{a}v^{a} + b^{a}
\end{equation}
We then concatenate the context vector from the text encoders ($c^{t}$) with the audio context vector ($c^{a}$) to obtain the multimodal context vector ($c$) for a response.

\begin{equation}
    c = concat(c^{t}, c^{a})
\end{equation}


\begin{table*}[htbp]
\adjustbox{max width=\textwidth}{%
\begin{tabular} {|c|c|c|c|c|c|c|c|c|c|c|c|c|c|}
    \hline
    \textbf{Model} & \multicolumn{2}{|c|}{\textbf{Prompt 1}} & \multicolumn{2}{|c|}{\textbf{Prompt 2}} & \multicolumn{2}{|c|}{\textbf{Prompt 3}} & \multicolumn{2}{|c|}{\textbf{Prompt 4}} & \multicolumn{2}{|c|}{\textbf{Prompt 5}} & \multicolumn{2}{|c|}{\textbf{Prompt 6}} & \textbf{Average}\\
    \hline
     & QWK & MSE
    & QWK & MSE
    & QWK & MSE
    & QWK & MSE
    & QWK & MSE
    & QWK & MSE & QWK \\
    \hline
    \multicolumn{14}{|c|}{\textbf{Baseline Modelling Strategy}} \\
    \hline
    BDLSTM & \textbf{0.4984} & 0.2208 & 0.3298 & 0.4923 & 0.4472 & 0.5847 & 0.4026 & 0.7944 & 0.3718 & 0.9868 & \textbf{0.3744} & \textbf{0.1534} & 0.4040 \\
    \hline
    BDLSTM+Attn & 0.5291 & \textbf{0.2284} & \textbf{0.3578} & 0.4784 & 0.5050 & 0.6715 & 0.5007 & 0.6277 & 0.5324 & \textbf{0.5277} & 0.4254 & 0.1881 & 0.4751 \\
    \hline
    BERT & \textcolor{mygreen}{\textbf{0.5443}} & \textcolor{mygreen}{\textbf{0.2166}} & 0.3566 & 0.4763 & 0.5148 & 0.6305 & 0.5302 & 0.5375 & 0.5494 & 0.5555 & 0.4333 & 0.1777 & 0.4881 \\
    \hline
    BDLSTM+Audio & 0.5059 & 0.2336 & 0.3266 & 0.4964 & 0.4732 & 0.5880 & 0.4888 & 0.5951 & 0.4809 & 0.6455 & 0.3922 & 0.1612 & 0.4446 \\
    \hline
    BDLSTM+Attn+Audio & \textbf{0.5173} & 0.2464 & 0.3402 & 0.5255 & 0.5390 & 0.5525 & 0.5286 & 0.5724 & 0.5394 & 0.5269 & 0.3859 & 0.2130 & 0.4751 \\
    \hline
    BERT+Audio & \textbf{0.5385} & 0.2230 & 0.3716 & 0.4704 & 0.5420 & 0.5246 & 0.5289 & 0.5253 & 0.5486 & 0.5239 & 0.4208 & \textbf{0.1820} & 0.4917 \\
    \hline

    \multicolumn{14}{|c|}{\textbf{One Stage Speaker-Conditioned Hierarchical Modelling Strategy}} \\
    \hline
    BDLSTM & \textbf{0.4984} & 0.2208 & \textbf{0.3428} & \textbf{0.4645} & 0.4390 & 0.5791 & 0.4450 & 0.6548 & 0.5056 & 0.5187 & 0.3398 & 0.2187 & 0.4284 \\
    \hline
    BDLSTM+Attn & 0.5291 & \textbf{0.2284} & 0.3458 & 0.5125 & 0.5334 & 0.5888 & 0.5330 & 0.6222 & 0.5460 & 0.5805 & 0.4216 & 0.1805 & 0.4848 \\
    \hline
    BERT & \textcolor{mygreen}{\textbf{0.5443}} & \textcolor{mygreen}{\textbf{0.2166}} & 0.3492 & 0.4569 & 0.5772 & 0.4423 & 0.5589 & \textbf{0.4479} & \textbf{0.5820} & \textbf{0.4486} & \textcolor{mygreen}{\textbf{0.4726}} & 0.1583 & 0.5140 \\
    \hline
    BDLSTM+Audio & 0.5059 & 0.2336 & \textbf{0.3447} & \textbf{0.4810} & 0.5075 & 0.5468 & 0.5258 & 0.5333 & \textbf{0.5281} & 0.4865 & \textbf{0.4147} & \textbf{0.1498} & 0.4711 \\
    \hline
    BDLSTM+Attn+Audio & \textbf{0.5173} & 0.2464 & \textcolor{mygreen}{\textbf{0.3776}} & 0.4481 & \textbf{0.5716} & 0.5213 & \textbf{0.5657} & 0.5142 & 0.5660 & 0.5241 & \textbf{0.4610} & \textbf{0.1732} & \textbf{0.5099} \\
    \hline
    BERT+Audio & \textbf{0.5385} & 0.2230 & \textbf{0.3746} & \textbf{0.4593} & 0.5684 & 0.4642 & 0.5601 & 0.4711 & 0.5770 & 0.4614 & 0.4256 & 0.1938 & 0.5074\\
    \hline

    \multicolumn{14}{|c|}{\textbf{Two Stage Speaker-Conditioned Hierarchical Modelling Strategy}} \\
    \hline
    BDLSTM & 0.4951 & \textbf{0.2194} & 0.3257 & 0.4833 & \textbf{0.5307} & \textbf{0.4826} & \textbf{0.5527} & \textcolor{mygreen}{\textbf{0.4256}} & \textbf{0.5279} & \textbf{0.4631} & 0.3040 & 0.2541 & \textbf{0.4560} \\ 
    \hline
    BDLSTM+Attn & \textbf{0.5350} & 0.2298 & 0.3563 & \textbf{0.4437} & \textbf{0.5575} & \textbf{0.5131} & \textbf{0.5592} & \textbf{0.5090} & \textbf{0.5585} & 0.5326 & \textbf{0.4348} & \textbf{0.1694} & \textbf{0.5002} \\ 
    \hline
    BERT & 0.5361 & 0.2243 & \textbf{0.3771} & \textbf{0.4451} & \textbf{0.5911} & \textbf{0.4319} & \textcolor{mygreen}{\textbf{0.5765}} & 0.4527 & 0.5693 & 0.5145 & 0.4701 & \textcolor{mygreen}{\textbf{0.1472}} & \textcolor{mygreen}{\textbf{0.5200}} \\ 
    \hline
    BDLSTM+Audio & \textbf{0.5422} & \textbf{0.2173} & 0.3407 & 0.5028 & \textbf{0.5732} & \textcolor{mygreen}{\textbf{0.4005}} & \textbf{0.5872} & \textbf{0.4353} & 0.5190 & \textbf{0.4857} & 0.3031 & 0.2592 & \textbf{0.4776} \\ 
    \hline
    BDLSTM+Attn+Audio & 0.5104 & \textbf{0.2443} & 0.3637 & \textcolor{mygreen}{\textbf{0.4332}} & 0.5669 & \textbf{0.5092} & 0.5614 & \textbf{0.5120} &\textbf{ 0.5663} & \textbf{0.4801} & 0.4212 & 0.1740 & 0.4983 \\ 
    \hline
    BERT+Audio & 0.5173 & \textbf{0.2216} & 0.3464 & 0.4607 & \textcolor{mygreen}{\textbf{0.5992}} & \textbf{0.4204} & \textbf{0.5727} & \textbf{0.4558} & \textcolor{mygreen}{\textbf{0.5927}} & \textcolor{mygreen}{\textbf{0.4475}} & \textbf{0.4552} & 0.1827 & \textbf{0.5139 } \\ 
    \hline
\end{tabular}}
\caption{\small \label{tab:results}Quadratic Weighted Kappa (QWK) score and Mean Squared Error (MSE) across prompts for different models. We present the results for the three modeling strategies (baseline, one stage speaker-conditioned hierarchical and two stage speaker-conditioned hierarchical) and 6 models (BDLSTM, BDLSTM+Attn, BERT, BDLSTM+Attn+wav2vec2.0, BERT+wav2vec2.0). Bold text represents best modeling technique for the model-prompt pair and \textcolor{mygreen}{green} text represents best model for each prompt. It is noteworthy that most \textcolor{mygreen}{green} and bold values lie in the proposed one-stage and two-stage speaker-conditioned hierarchical models.}
\end{table*}

\subsection{Modeling Strategies}
\label{sec:Modeling strategies}
We try out three modeling strategies: baseline strategy without any speaker-conditioning (Fig.~\ref{fig:baseline}), one stage speaker-conditioned hierarchical modeling (Fig.~\ref{fig:hierarchical}), two stage speaker-conditioned hierarchical modeling (Fig.~\ref{fig:two stage hierarchical model}). In the speaker-conditioned hierarchical modeling strategies, we provide other responses ($r_{ij}$ for different values of $j$) of the same speaker ($s_i$) on other prompts ($p_j$) as context ($c_{ij}$) to the model ($m_j$) to assist its predictions ($o_{ij}$). Since a response $r_{ij}$ is dependent on speaker $s_i$ and prompt $p_j$, we hypothesise that providing the model with a test taker's context can be a crucial element in understanding his/her speech and making a more informed prediction.

\textbf{Baseline Strategy:} 
This is given by the Fig.~\ref{fig:baseline} where we have 6 models ($m_j,  j\in\{1..6\}$), one for each prompt ($p_j,  j\in\{1..6\}$). The models are conditioned only on text transcripts ($T_{ij}$) or a combination of text and audio ($concat(T_{ij}, A_{ij})$). The text transcripts and audio are encoded by the text and audio encoders ($E_j, j\in\{1..6\}$) explained in the previous section (\S\ref{sec:models}). We obtain the context vector $c_{ij}$ from encoder $E_{j}$ for the response of speaker $s_{i}$ on prompt $p_{j}$. We then pass this context vector through a fully connected layer to obtain the final score $o_{ij}$.
\begin{equation}
    o_{ij} = W_{m_{j}}c_{ij} + b_{m_{j}}
\end{equation}
The models are trained independently of each other using weighted MSE loss until convergence.

\textbf{One Stage Speaker-Conditioned Hierarchical Model:} As we show in Fig.~\ref{fig:hierarchical}, in this approach, we condition prompt $p_j$'s scoring model $m_j$ on all the inputs ($T_{ik}, A_{ik}$) of prompts $p_k$ s.t. $k<j$.
The context of a response is the context vector extracted from the encoder model (\textit{e.g.} final hidden state in BDLSTM).
\begin{equation}
    c_{ij} = E_{j}(T_{ij}, A_{ij})
\end{equation}
where $c_{ij}$ is the context vector of a model on the $j^{th}$ prompt for the $i^{th}$ speaker ($s_{i}$). We define the new context vector $c'_{ij}$ as:
\begin{equation}
    c'_{ij} = concat(c_{i1}, c_{i2}, ..., c{ij})
\end{equation}
It is the concatenation of context vectors from the previous prompts and the context vector from the current prompt $j$ for a particular speaker $s_{i}$. We then pass this new context vector $c'_{ij}$ through a fully connected layer to obtain a score for the sample $o_{ij}$.

\begin{equation}
    o_{ij} = W_{m_{j}}c'_{ij} + b_{m_{j}}
\end{equation}

We train our models sequentially from prompt 1 to prompt 6, and store the context vector for every speaker after the training completes. While training the model on the $j^{th}$ prompt, the context vectors $c_{i1}$ to $c_{ij-1}$ are stored vectors obtained from previously trained models (Fig.~\ref{fig:hierarchical}). We perform this experiment on all of our models, \textit{i.e.}, BDLSTM, BDLSTM+Attention, BERT, and text encoders with wav2vec2.0.


\textbf{Two Stage Speaker-Conditioned Hierarchical Modeling:}
In this approach (Fig.~\ref{fig:two stage hierarchical model}), we assist the training of our model by providing it the context of all the responses of a speaker from the baseline models as shown in Fig.~\ref{fig:two stage hierarchical model}. As modeling non-native users is a harder problem \cite{bejar2017threats,flek2020returning,shah2021all}, our model needs more user specific context. The context of response is the context vector extracted from the baseline model (\textit{e.g.} final hidden state in BDLSTM) that is then passed through the fully connected layer to score the sample. 
\begin{equation}
    b_{ij} = E_{j}(T_{ij}, A_{ij})
\end{equation}
Let $b_{ij}$ be the stored context vector of a trained baseline model trained on the $j^{th}$ prompt for the $i^{th}$ speaker ($s_{i}$). Then the global context vector for $s_{i}$ be defined as
\begin{equation}
    v_{i}= concat(b_{ij,\forall j \neq i}^{6})
\end{equation}
We define the new context vector $c'_{ij}$ as
\begin{equation}
    c'_{ij} = concat(v_{i}, c_{ij})
\end{equation}
as the concatenation of the global baseline context vector $v_{i}$ and the context vector from the current prompt $c_{ij}$ for a particular speaker.
We then pass this new context vector ($c'_{ij}$) through a fully connected layer to obtain a score for the sample $o_{ij}$.

\begin{equation}
     o_{ij} = W_{m_{j}}c'_{ij} + b_{m_{j}}
\end{equation}

We perform this experiment on all of our models, \textit{i.e.}, BDLSTM, BDLSTM+Attention, BERT, and text encoders with wav2vec2.0 (Fig.~\ref{fig:two stage hierarchical model}).

\begin{figure*}[h]
    \centering
    \subfloat[][Number of speakers for whom output scores of at least $k$ prompts are predicted correctly.]{
    \includegraphics[width=0.75\columnwidth]{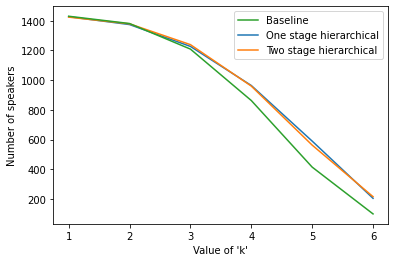}
    \label{fig:atleast_k_participants}}
    \qquad
    \subfloat[][Relative improvement over baseline (in \%) in the number of speakers for whom output scores of at least $k$ prompts are predicted correctly.]{
    \includegraphics[width=0.75\columnwidth]{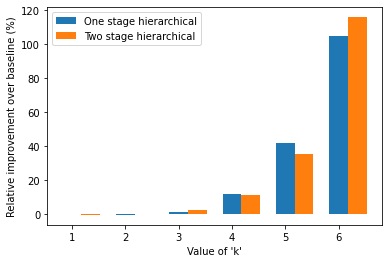}
    \label{fig:atleast_k_relative}}
    \caption{\small Speaker level accuracy comparison in the three modeling strategies. For generating these plots, we use the results from BERT model.}
    \label{fig:atleast_k}
\end{figure*}

\subsection{Training}
\label{sec:training}
We use the weighted mean square loss as our loss function defined as:
\begin{equation}
    WMSE(y_{true}, y_{pred}) = \frac{1}{N}\sum_{i=1}^{N}(y^{i}_{true} - y^{i}_{pred})^{2}w_{y^{i}_{true}}
\end{equation}

where $y^{i}_{true}$, $y^{i}_{pred}$ are the scaled ground truth, and predicted grade, respectively. $w_{y_{true}}$ is the weight of the ground truth class. We use the weighted mean squared loss due to class imbalance in the training data as shown in Table~\ref{tab:dataset}. We use the Adam \cite{adam} optimizer to minimize our loss over the training data. To prevent overfitting, we train the model with early stopping on the weighted validation MSE loss. We also use the ReduceLROnPlateau scheduler as provided in PyTorch \cite{pytorch} to reduce the learning rate when the weighted validation MSE loss does not improve. 

\subsection{Evaluation Metric}
\label{sec:evaluation metric}
Similar to previous studies
\cite{qian2018prompt,wang2018automatic,wang2021automated,grover2020multi,kumar2019get,parekh2020my,kumar2020calling}, we have used the quadratic weighted kappa (QWK) as our evaluation metric. It measures the level of agreement between two sets of ratings. The metric outputs values between 0 (random agreement between ratings) and 1 (complete agreement between ratings). In cases where the agreement is less than expected by chance, the score can be negative too. To calculate the QWK score, firstly we calculate the confusion matrix $O$ of size $C$x$C$, where $C$ is the total number of classes. $O_{ij}$ represents the number of samples with true label $i$ and predicted label $j$. The matrix is normalised to have sum equal 1. We calculate the weight matrix $W$ of size $C$x$C$, where $W_{ij} = (i - j)^{2}/(C - 1)^{2}$. The weight matrix penalises predictions that are further away from their ground truth more harshly than the ones that are closer to the ground truth. Then we create the histogram matrix of expected grades $E$ of size $C$x$C$, which is calculated by taking the outer product of histograms of ground truth and predicted labels. We normalise this matrix such that the sum equals 1. The QWK is calculated as:
\begin{equation}
QWK = 1 - (\sum_{i,j}W_{ij}O_{ij})/(\sum_{i,j}W_{ij}E_{ij})
\end{equation}

Apart from QWK metric, following the recommendations of other studies \cite{madnani2018automated,kumar2020calling}, we analyze the results more comprehensively by calculating speaker-level accuracy, performance on near decision-boundary samples and high-biased samples. In order to gain more intuition, we perform attributions over both modalities and also show information sharing strength across prompts. The results of all these analyses are presented in the next section.


\section{Results}
\label{sec:results}
Here we present the quantitative and qualitative results for the different models and modeling strategies. 

\subsection{Quantitative Analysis}
\label{sec:quantitative analysis}
\textbf{QWK Score:} In Table \ref{tab:results}, we compare the two proposed speaker-conditioned hierarchical approaches with baseline strategies on various models. It can be inferred from the table that \textit{speaker-conditioned hierarchical modeling consistently improves performance across all the models on all prompts}. This result supports our hypothesis that speaker-specific cues assist the model in making better informed predictions. In table \ref{tab:results}, we also report the average QWK for each model which we define as the average of promptwise QWK values for a model. The mean improvement in average QWK across all models is 4.97\% (maximum = 7.32\%, minimum = 2.05\%) and 6.92\% (maximum = 12.86\%, minimum = 4.51\%)  in one stage and two stage hierarchical models compared to their baseline counterparts. 

Similar to average QWK, we also compute average MSE for each model. As compared to the baseline, we observe a mean decrease in average MSE by 10.03\% and 15.21\% in one stage and two stage hierarchichal models, respectively.


We see that the two stage speaker-conditioned hierarchical BERT model performs better than the other models in three out of six prompts. For the other three, the one stage speaker-conditioned hierarchical BERT model performs better. This result strongly indicates that conditioning on speakers has improved the results.

Table~\ref{tab:results} also contains results for experiments with text and audio features. Similar trends as text-only models are also observed here. First, we see an improvement in  performance of 61.11\% multimodal models as compared to the text-only models. Our one stage and two stage speaker-conditioned hierarchical BDLSTM+Attn+Wav2Vec2.0 model beats the  BERT text-only model. This shows that our hypothesis of audio-specific features not being captured within text transcriptions was true and including wav2vec2.0's features helped improve the performance.
Speaker-conditioning the models further improves the performance by 5.9\% as compared to the multi-modal baseline models. 




\textbf{Speaker Level Accuracy:} Since our hypothesis is that including speaker-level information from previous models should improve the performance of subsequent models, we also show improvements in speaker-level accuracy. For this, we first compare the mean number of prompts correctly classified for each speaker by the three modeling strategies, \textit{i.e.}, $\frac{1}{N}\sum_{i=1,j=1}^{i=N,j=6} I(y_{true,i,j}=y_{pred,i,j})$ where $y_{pred,i,j}$, $y_{true,i,j}$ are the predicted output and the ground truth for $i^{th}$ speaker on $j^{th}$ prompt. We found that the mean number of correctly classified prompts for one stage hierarchical and two stage hierarchical models are 4.018 and 4.014, respectively, which is 7.4\% better than the baseline mean of 3.740.

To strengthen our claims further for speaker level accuracy, we compare the number of speakers for whom responses on at least $k$ prompts are predicted correctly for the three modeling strategies. Concretely, we measure, $Count_{i=1}^{N}(I(\sum_{j=1}^{6}I(y_{true,i,j}=y_{pred,i,j}) \geq k))$ for different values of $k$.  We present the results in Fig.~\ref{fig:atleast_k_participants}, where we see that both the one-stage hierarchical model and the two stage hierarchical model outperform the baseline for all the values of $k$. In Fig. \ref{fig:atleast_k_relative}, we can see the significance of this improvement as compared to the baseline model. As the value of k increases, the relative improvements become more significant. This observation suggests that speaker-specific cues used in speaker-conditioned modeling help our model increase speaker level accuracy.
\begin{figure*}[t!]
    \centering
    \includegraphics[width=15cm]{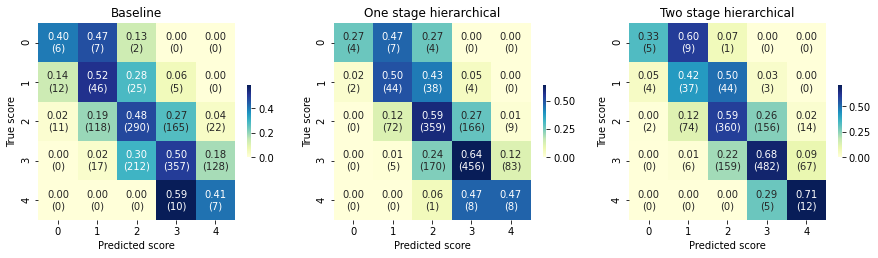}
    \caption{\small Heatmap on prompt 4, each element $e_{ij}$ depicts percentage (number) of samples predicted in score $j$ having ground truth score $i$. The blue patches clearly indicate that there are lesser samples satisfying $|y_{true}-y_{pred}| \geq2$ in case of our proposed models when compared to the baseline models}
    \label{fig:bias_precision}
\end{figure*}

\textbf{High Bias Samples:} We set out to see if conditioning on speakers makes the models avoid errors where the model's predictions are far from the ground truth ($|y_{true}-y_{pred}| \geq2$). Based on our hypothesis, we expect that given more speaker level information, the models should be able to perform better on such samples. Consequently, we analyze samples which were predicted incorrectly by more than 2 score levels. As compared to baseline models, we observe upto \textbf{60\%} and \textbf{55\%} reduction in high-bias errors in one stage and two stage hierarchical models, respectively. Results obtained on prompt 4 are presented in Fig.~\ref{fig:bias_precision}. This shows that giving more speaker-specific context reduces errors on high-bias samples.

\textbf{Bias Due To Performance On Other Prompts:}
Since with speaker conditioning, we are giving more speaker context to the models, models might become unfairly biased towards a speaker's previous (good or bad) performance in some prior prompt. For instance, if a speaker performs poorly (well) on some prior prompt but better (poor) on the current prompt, the speaker should not be unfairly penalized (awarded) because of his previous performance. 

We analyse the models' predictions for those samples on which the speakers have a high predicted score on prompt $i-1$ and low ground-truth scores on prompt $i$. For the other case (where predicted score on prompt i - 1 is low and ground truth on prompt i is high), the number of samples were too less to obtain any statistical inference. We find that one stage and two stage hierarchical models tend to give a higher score than the ground truth score relative to the baseline in 3.5\%, 7\% samples, respectively. Although this was not consistent across all prompts. Moreover, the average percentage of correctly predicted samples in the one stage and two stage hierarchical models were higher by 10.30\% (maximum = 28.00\%, minimum = 4.10\%) and 10.05\% (maximum = 36.34\%, minimum = 4.43\%), respectively, than the corresponding baseline models. Therefore, despite the increase in predictions observed in 3 out of 6 prompts, the percentage increase in correctly predicted samples is much more significant. This shows that the model extracts speaker-specific features and rather than being biased by it, the model uses it to improve its performance on the current prompt and be more precise. We leave the correction of marginal score change on 3 out of 6 prompts as future work.


\begin{figure*}[h]
    \centering
    \subfloat[][Accuracy]{
    \includegraphics[width=0.75\columnwidth]{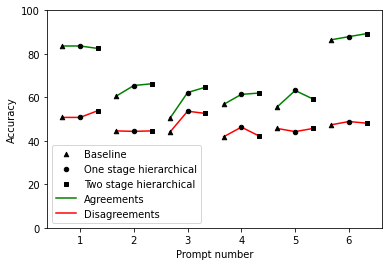}
    \label{fig:human_annotators_accuracy}}
    \qquad
    \subfloat[][QWK]{
    \includegraphics[width=0.75\columnwidth]{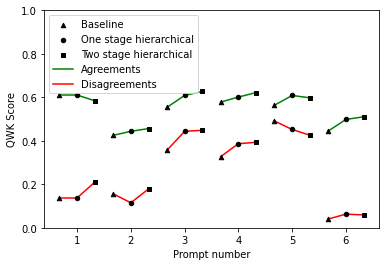}
    \label{fig:human_annotators_qwk}}

\caption{\small (a) Accuracy and (b) QWK on two types of samples: (i) Where the two expert raters agree (ii) Where the two expert raters disagree}
\label{fig:human_annotators}
\end{figure*}

\textbf{Performance On Samples Near Decision Boundary:} 
We define near decision boundary samples as the ones where the two raters disagree on speech scores. Fig.~\ref{fig:human_annotators} presents the results for both types of samples, \textit{i.e.}, where human raters (i) agree and (ii) disagree with each other. We observe that conditioning on speakers improves the performance on both types of samples. Average accuracy increase is 14.91\% and 6.33\% in agreements and disagreements for prompts 2, 3, and 5 over the baseline models. Further, we observe that all modelling strategies have a lower accuracy and QWK for samples where the two raters disagree. This is expected since these samples are hard for both humans and models and can be considered to lie near the model's decision boundary.
\subsection{Qualitative Analysis}
\label{sec:Qualitative Analysis}

In Section~\ref{sec:quantitative analysis}, we see an improvement in the QWK score. In order to further probe how our technique helps in improving model performance, we use an attribution method to analyze models' predictions with respect to their inputs. We employ the method of integrated gradients \cite{sundararajan2017axiomatic} for this purpose:

Given an input $x$ and a baseline
$b$\footnote{Defined as an input containing absence of cause for the output of a model; also called neutral input \cite{shrikumar2016not,sundararajan2017axiomatic}.}, the integrated gradient along the $i^{th}$ dimension is defined as:
\begin{equation}
\label{eqn:intgrad}
IG_i(x,b) = (x_i-b_i)\int_{\alpha=0}^{1} \frac{\partial F(b + \alpha(x-b))}{\partial x_i  }~d\alpha
\end{equation}
where $\frac{\partial F(x)}{\partial x_i}$ represents the gradient of $F$ along the $i^{th}$ dimension of  $x$. We choose the baseline as empty input (all 0s) since an empty speech sample should get a score of 0 as per the scoring rubrics. For the first experiment on prompt-wise attribution, we took prompt-wise response embeddings as the dimension of attribution. For the second experiment on finding the contribution of modalities, we took text and audio modalities across all prompts, as the two dimensions for finding attributions.

Integrated gradients have been successfully used in the past for various NLP tasks \cite{shrikumar2016not,parekh2020my,mudrakarta2018did}. To calculate these attributions, we used the Captum \cite{captumai} library and aggregated all attributions of the test set to retain a global view of the importance of different embeddings.

\begin{figure}[h]
    \centering
    \includegraphics[width=7cm]{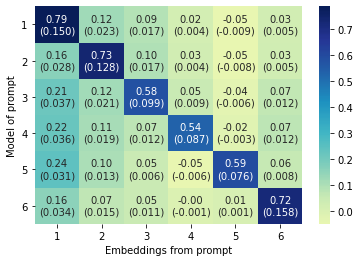}
    \caption{\small Heatmap for promptwise attributions calculated using BERT two stage speaker-conditioned hierarchical model. Here, each element $e_{ij}$ depicts contibution (attribution value) of prompt $j's$ embeddings in predicting the output score of prompt ${i}$ }
    \label{fig:attributions_two_stage}
\end{figure}

\textbf{Prompt-wise Attribution}: To assess information sharing strength across prompts, we compute prompt-wise attributions on the two stage hierarchical models. The results for this are shown in Fig.~\ref{fig:attributions_two_stage}. Here we observe that the most crucial embedding in predicting each prompt is the embedding from the prompt itself, which is expected since this embedding contains most of the context related to the response of that prompt. Additionally, we can see that there are many lighter patches in Fig.~\ref{fig:attributions_two_stage}. These patches represent the importance of embeddings from other prompts in the current model's predictions. As can be seen in the heatmap, embeddings from prompts 1 and 2 were the most important and contributed significantly in predicting other prompts. This validates our hypothesis that providing a model with rich user specific context via hierarchical modeling can be useful for improving its performance.

\begin{figure}[h]
    \centering
    \includegraphics[width=6cm]{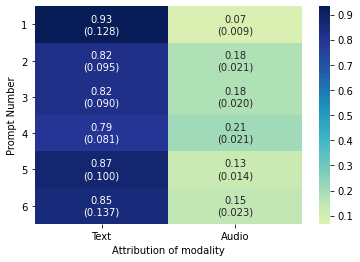}
    \caption{\small Attribution value of each modality in predicting other prompts calculated using BERT+wav2vec2.0 two stage speaker-conditioned hierarchical model.}
    \label{fig:attributions_multimodal}
\end{figure}

\textbf{Contribution Of Modalities}: To understand the importance of text and audio individually in predicting output scores of different prompts, we calculated attribution scores over them (see Fig.~\ref{fig:attributions_multimodal}). We observe that attribution scores are lower for audio as compared to text for all prompts. This demonstrates that our text encoders were able to capture most of the information required for predicting scores and audio embeddings fulfilled the shortcomings of the text encoding process (like pronunciation and prosody which cannot be directly extracted from text transcripts).

We also observe that the relative contribution of audio was the highest for prompts 3 and 4 which are incidentally the most difficult prompts (Table~\ref{tab:dataset}). This is expected since for more difficult prompts, apart from content, the presentation also plays an important role. On the other hand, prompts 1  and 6 have the highest contribution of text. These prompts had the lowest difficulty level (B1) and hence content plays a dominant role in the score predictions for them. We observed similar plots for both one stage speaker-conditioned and baseline models.


\section{Conclusion}
\label{sec:conclusion}
In this paper, we have investigated end-to-end deep learning based models for the automated speech scoring task for L2 English speakers. We show that speaker-conditioned hierarchical models perform better than their baseline counterparts. Results indicate that providing the model with additional speaker-specific context helps the model in better understanding the speaker and making a more informed prediction. These models are more precise and less biased than their baseline counterparts. We also show that including audio based features increases models' expressivity and generalisability, which improves the overall performance. In future work, we wish to explore the effectiveness of our technique on AES, off-topic response detection, and individual speech components such as dysfluency and pronunciation.
\\\\\textbf{Acknowledgements:} Yaman Kumar is the recipient of Google PhD fellowship and would like to thank Google for supporting him. We would also like to thank Second Language Testing Inc. (SLTI) for providing us with data. Rajiv Ratn Shah is partly supported by the Infosys Center for AI and Center for Design and New Media at IIIT-Delhi and would like to thank them.


\newpage
\bibliographystyle{ACM-Reference-Format}
\bibliography{sample-base}


\begin{thebibliography}{76}


\ifx \showCODEN    \undefined \def \showCODEN     #1{\unskip}     \fi
\ifx \showDOI      \undefined \def \showDOI       #1{#1}\fi
\ifx \showISBNx    \undefined \def \showISBNx     #1{\unskip}     \fi
\ifx \showISBNxiii \undefined \def \showISBNxiii  #1{\unskip}     \fi
\ifx \showISSN     \undefined \def \showISSN      #1{\unskip}     \fi
\ifx \showLCCN     \undefined \def \showLCCN      #1{\unskip}     \fi
\ifx \shownote     \undefined \def \shownote      #1{#1}          \fi
\ifx \showarticletitle \undefined \def \showarticletitle #1{#1}   \fi
\ifx \showURL      \undefined \def \showURL       {\relax}        \fi
\providecommand\bibfield[2]{#2}
\providecommand\bibinfo[2]{#2}
\providecommand\natexlab[1]{#1}
\providecommand\showeprint[2][]{arXiv:#2}

\bibitem[\protect\citeauthoryear{Amodei, Ananthanarayanan, Anubhai, Bai,
  Battenberg, Case, Casper, Catanzaro, Cheng, Chen, Chen, Chen, Chen,
  Chrzanowski, Coates, Diamos, Ding, Du, Elsen, Engel, Fang, Fan, Fougner, Gao,
  Gong, Hannun, Han, Johannes, Jiang, Ju, Jun, LeGresley, Lin, Liu, Liu, Li,
  Li, Ma, Narang, Ng, Ozair, Peng, Prenger, Qian, Quan, Raiman, Rao, Satheesh,
  Seetapun, Sengupta, Srinet, Sriram, Tang, Tang, Wang, Wang, Wang, Wang, Wang,
  Wang, Wu, Wei, Xiao, Xie, Xie, Yogatama, Yuan, Zhan, and Zhu}{Amodei
  et~al\mbox{.}}{2016}]%
        {deepspeech2}
\bibfield{author}{\bibinfo{person}{Dario Amodei}, \bibinfo{person}{Sundaram
  Ananthanarayanan}, \bibinfo{person}{Rishita Anubhai},
  \bibinfo{person}{Jingliang Bai}, \bibinfo{person}{Eric Battenberg},
  \bibinfo{person}{Carl Case}, \bibinfo{person}{Jared Casper},
  \bibinfo{person}{Bryan Catanzaro}, \bibinfo{person}{Qiang Cheng},
  \bibinfo{person}{Guoliang Chen}, \bibinfo{person}{Jie Chen},
  \bibinfo{person}{Jingdong Chen}, \bibinfo{person}{Zhijie Chen},
  \bibinfo{person}{Mike Chrzanowski}, \bibinfo{person}{Adam Coates},
  \bibinfo{person}{Greg Diamos}, \bibinfo{person}{Ke Ding},
  \bibinfo{person}{Niandong Du}, \bibinfo{person}{Erich Elsen},
  \bibinfo{person}{Jesse Engel}, \bibinfo{person}{Weiwei Fang},
  \bibinfo{person}{Linxi Fan}, \bibinfo{person}{Christopher Fougner},
  \bibinfo{person}{Liang Gao}, \bibinfo{person}{Caixia Gong},
  \bibinfo{person}{Awni Hannun}, \bibinfo{person}{Tony Han},
  \bibinfo{person}{Lappi~Vaino Johannes}, \bibinfo{person}{Bing Jiang},
  \bibinfo{person}{Cai Ju}, \bibinfo{person}{Billy Jun},
  \bibinfo{person}{Patrick LeGresley}, \bibinfo{person}{Libby Lin},
  \bibinfo{person}{Junjie Liu}, \bibinfo{person}{Yang Liu},
  \bibinfo{person}{Weigao Li}, \bibinfo{person}{Xiangang Li},
  \bibinfo{person}{Dongpeng Ma}, \bibinfo{person}{Sharan Narang},
  \bibinfo{person}{Andrew Ng}, \bibinfo{person}{Sherjil Ozair},
  \bibinfo{person}{Yiping Peng}, \bibinfo{person}{Ryan Prenger},
  \bibinfo{person}{Sheng Qian}, \bibinfo{person}{Zongfeng Quan},
  \bibinfo{person}{Jonathan Raiman}, \bibinfo{person}{Vinay Rao},
  \bibinfo{person}{Sanjeev Satheesh}, \bibinfo{person}{David Seetapun},
  \bibinfo{person}{Shubho Sengupta}, \bibinfo{person}{Kavya Srinet},
  \bibinfo{person}{Anuroop Sriram}, \bibinfo{person}{Haiyuan Tang},
  \bibinfo{person}{Liliang Tang}, \bibinfo{person}{Chong Wang},
  \bibinfo{person}{Jidong Wang}, \bibinfo{person}{Kaifu Wang},
  \bibinfo{person}{Yi Wang}, \bibinfo{person}{Zhijian Wang},
  \bibinfo{person}{Zhiqian Wang}, \bibinfo{person}{Shuang Wu},
  \bibinfo{person}{Likai Wei}, \bibinfo{person}{Bo Xiao}, \bibinfo{person}{Wen
  Xie}, \bibinfo{person}{Yan Xie}, \bibinfo{person}{Dani Yogatama},
  \bibinfo{person}{Bin Yuan}, \bibinfo{person}{Jun Zhan}, {and}
  \bibinfo{person}{Zhenyao Zhu}.} \bibinfo{year}{2016}\natexlab{}.
\newblock \showarticletitle{Deep Speech 2: End-to-End Speech Recognition in
  English and Mandarin}. In \bibinfo{booktitle}{\emph{Proceedings of the 33rd
  International Conference on International Conference on Machine Learning -
  Volume 48}} (New York, NY, USA) \emph{(\bibinfo{series}{ICML'16})}.
  \bibinfo{publisher}{JMLR.org}, \bibinfo{pages}{173–182}.
\newblock


\bibitem[\protect\citeauthoryear{Ardila, Branson, Davis, Henretty, Kohler,
  Meyer, Morais, Saunders, Tyers, and Weber}{Ardila et~al\mbox{.}}{2020}]%
        {commonvoice}
\bibfield{author}{\bibinfo{person}{Rosana Ardila}, \bibinfo{person}{Megan
  Branson}, \bibinfo{person}{Kelly Davis}, \bibinfo{person}{Michael Henretty},
  \bibinfo{person}{Michael Kohler}, \bibinfo{person}{Josh Meyer},
  \bibinfo{person}{Reuben Morais}, \bibinfo{person}{Lindsay Saunders},
  \bibinfo{person}{Francis~M. Tyers}, {and} \bibinfo{person}{Gregor Weber}.}
  \bibinfo{year}{2020}\natexlab{}.
\newblock \showarticletitle{Common Voice: A Massively-Multilingual Speech
  Corpus}. In \bibinfo{booktitle}{\emph{LREC}}.
\newblock


\bibitem[\protect\citeauthoryear{Association et~al\mbox{.}}{Association
  et~al\mbox{.}}{2014}]%
        {educational2014snapshot}
\bibfield{author}{\bibinfo{person}{Educational~Testing Association}
  {et~al\mbox{.}}} \bibinfo{year}{2014}\natexlab{}.
\newblock \bibinfo{title}{A snapshot of the individuals who took the GRE
  revised general test}.
\newblock
\newblock


\bibitem[\protect\citeauthoryear{Baevski, Zhou, Mohamed, and Auli}{Baevski
  et~al\mbox{.}}{2020}]%
        {wav2vec2}
\bibfield{author}{\bibinfo{person}{Alexei Baevski}, \bibinfo{person}{Henry
  Zhou}, \bibinfo{person}{Abdelrahman Mohamed}, {and} \bibinfo{person}{Michael
  Auli}.} \bibinfo{year}{2020}\natexlab{}.
\newblock \showarticletitle{wav2vec 2.0: {A} Framework for Self-Supervised
  Learning of Speech Representations}.
\newblock \bibinfo{journal}{\emph{CoRR}}  \bibinfo{volume}{abs/2006.11477}
  (\bibinfo{year}{2020}).
\newblock
\showeprint[arxiv]{2006.11477}
\urldef\tempurl%
\url{https://arxiv.org/abs/2006.11477}
\showURL{%
\tempurl}


\bibitem[\protect\citeauthoryear{Bamdev, Grover, Singla, Shah, Vafaee, and
  Hama}{Bamdev et~al\mbox{.}}{2021}]%
        {bamdevAutomated}
\bibfield{author}{\bibinfo{person}{Pakhi Bamdev}, \bibinfo{person}{Manraj~Singh
  Grover}, \bibinfo{person}{Yaman~Kumar Singla}, \bibinfo{person}{Rajiv~Ratn
  Shah}, \bibinfo{person}{Payman Vafaee}, {and} \bibinfo{person}{Mika Hama}.}
  \bibinfo{year}{To appear in 2021}\natexlab{}.
\newblock \showarticletitle{Automated Speech Scoring System Under The Lens:
  Evaluating and interpreting the linguistic cues for language proficiency}.
\newblock \bibinfo{journal}{\emph{International Journal of Artificial
  Intelligence in Education}} (\bibinfo{year}{To appear in 2021}).
\newblock


\bibitem[\protect\citeauthoryear{Bejar}{Bejar}{2017}]%
        {bejar2017threats}
\bibfield{author}{\bibinfo{person}{Isaac~I Bejar}.}
  \bibinfo{year}{2017}\natexlab{}.
\newblock \showarticletitle{Threats to score meaning in automated scoring}.
\newblock \bibinfo{journal}{\emph{Validation of score meaning for the next
  generation of assessments: The use of response processes}}
  (\bibinfo{year}{2017}), \bibinfo{pages}{75--84}.
\newblock


\bibitem[\protect\citeauthoryear{Biewald}{Biewald}{2020}]%
        {wandb}
\bibfield{author}{\bibinfo{person}{Lukas Biewald}.}
  \bibinfo{year}{2020}\natexlab{}.
\newblock \bibinfo{title}{Experiment Tracking with Weights and Biases}.
\newblock
\newblock
\urldef\tempurl%
\url{https://www.wandb.com/}
\showURL{%
\tempurl}
\newblock
\shownote{Software available from wandb.com.}


\bibitem[\protect\citeauthoryear{Boersma and Weenink}{Boersma and
  Weenink}{2009}]%
        {Boersma2009praat}
\bibfield{author}{\bibinfo{person}{Paul Boersma} {and} \bibinfo{person}{David
  Weenink}.} \bibinfo{year}{2009}\natexlab{}.
\newblock \bibinfo{title}{Praat: doing phonetics by computer (Version 5.1.13)}.
\newblock
\newblock
\urldef\tempurl%
\url{http://www.praat.org}
\showURL{%
\tempurl}


\bibitem[\protect\citeauthoryear{Broeder and Martyniuk}{Broeder and
  Martyniuk}{2008}]%
        {cefr}
\bibfield{author}{\bibinfo{person}{Peter Broeder} {and}
  \bibinfo{person}{Waldemar Martyniuk}.} \bibinfo{year}{2008}\natexlab{}.
\newblock \bibinfo{booktitle}{\emph{Language Education in Europe: The Common
  European Framework of Reference}}.
\newblock \bibinfo{publisher}{Springer US}, \bibinfo{address}{Boston, MA},
  \bibinfo{pages}{1305--1322}.
\newblock
\showISBNx{978-0-387-30424-3}
\urldef\tempurl%
\url{https://doi.org/10.1007/978-0-387-30424-3_100}
\showDOI{\tempurl}


\bibitem[\protect\citeauthoryear{Chen, Tao, Ghaffarzadegan, and Qian}{Chen
  et~al\mbox{.}}{2018a}]%
        {chen2018end}
\bibfield{author}{\bibinfo{person}{Lei Chen}, \bibinfo{person}{Jidong Tao},
  \bibinfo{person}{Shabnam Ghaffarzadegan}, {and} \bibinfo{person}{Yao Qian}.}
  \bibinfo{year}{2018}\natexlab{a}.
\newblock \showarticletitle{End-to-end neural network based automated speech
  scoring}. In \bibinfo{booktitle}{\emph{2018 IEEE International Conference on
  Acoustics, Speech and Signal Processing (ICASSP)}}. IEEE,
  \bibinfo{pages}{6234--6238}.
\newblock


\bibitem[\protect\citeauthoryear{Chen, Tao, Ghaffarzadegan, and Qian}{Chen
  et~al\mbox{.}}{2018b}]%
        {Chen2018EndtoEndNN}
\bibfield{author}{\bibinfo{person}{Lei Chen}, \bibinfo{person}{Jidong Tao},
  \bibinfo{person}{Shabnam Ghaffarzadegan}, {and} \bibinfo{person}{Yao Qian}.}
  \bibinfo{year}{2018}\natexlab{b}.
\newblock \showarticletitle{End-to-End Neural Network Based Automated Speech
  Scoring}. In \bibinfo{booktitle}{\emph{2018 IEEE International Conference on
  Acoustics, Speech and Signal Processing (ICASSP)}}.
  \bibinfo{pages}{6234--6238}.
\newblock
\urldef\tempurl%
\url{https://doi.org/10.1109/ICASSP.2018.8462562}
\showDOI{\tempurl}


\bibitem[\protect\citeauthoryear{Devlin, Chang, Lee, and Toutanova}{Devlin
  et~al\mbox{.}}{2019}]%
        {bert}
\bibfield{author}{\bibinfo{person}{Jacob Devlin}, \bibinfo{person}{Ming-Wei
  Chang}, \bibinfo{person}{Kenton Lee}, {and} \bibinfo{person}{Kristina
  Toutanova}.} \bibinfo{year}{2019}\natexlab{}.
\newblock \showarticletitle{{BERT}: Pre-training of Deep Bidirectional
  Transformers for Language Understanding}. In
  \bibinfo{booktitle}{\emph{Proceedings of the 2019 Conference of the North
  {A}merican Chapter of the Association for Computational Linguistics: Human
  Language Technologies, Volume 1 (Long and Short Papers)}}.
  \bibinfo{publisher}{Association for Computational Linguistics},
  \bibinfo{address}{Minneapolis, Minnesota}, \bibinfo{pages}{4171--4186}.
\newblock
\urldef\tempurl%
\url{https://doi.org/10.18653/v1/N19-1423}
\showDOI{\tempurl}


\bibitem[\protect\citeauthoryear{Ding, Wang, Chang, Wan, and Moreno}{Ding
  et~al\mbox{.}}{2019}]%
        {ding2019personal}
\bibfield{author}{\bibinfo{person}{Shaojin Ding}, \bibinfo{person}{Quan Wang},
  \bibinfo{person}{Shuo-yiin Chang}, \bibinfo{person}{Li Wan}, {and}
  \bibinfo{person}{Ignacio~Lopez Moreno}.} \bibinfo{year}{2019}\natexlab{}.
\newblock \showarticletitle{Personal VAD: Speaker-conditioned voice activity
  detection}.
\newblock \bibinfo{journal}{\emph{arXiv preprint arXiv:1908.04284}}
  (\bibinfo{year}{2019}).
\newblock


\bibitem[\protect\citeauthoryear{Dong, Zhang, and Yang}{Dong
  et~al\mbox{.}}{2017}]%
        {dong-etal-2017-attention}
\bibfield{author}{\bibinfo{person}{Fei Dong}, \bibinfo{person}{Yue Zhang},
  {and} \bibinfo{person}{Jie Yang}.} \bibinfo{year}{2017}\natexlab{}.
\newblock \showarticletitle{Attention-based Recurrent Convolutional Neural
  Network for Automatic Essay Scoring}. In
  \bibinfo{booktitle}{\emph{Proceedings of the 21st Conference on Computational
  Natural Language Learning ({C}o{NLL} 2017)}}. \bibinfo{publisher}{Association
  for Computational Linguistics}, \bibinfo{address}{Vancouver, Canada},
  \bibinfo{pages}{153--162}.
\newblock
\urldef\tempurl%
\url{https://doi.org/10.18653/v1/K17-1017}
\showDOI{\tempurl}


\bibitem[\protect\citeauthoryear{Eskenazi}{Eskenazi}{2009}]%
        {ESKENAZI2009832}
\bibfield{author}{\bibinfo{person}{Maxine Eskenazi}.}
  \bibinfo{year}{2009}\natexlab{}.
\newblock \showarticletitle{An overview of spoken language technology for
  education}.
\newblock \bibinfo{journal}{\emph{Speech Communication}} \bibinfo{volume}{51},
  \bibinfo{number}{10} (\bibinfo{year}{2009}), \bibinfo{pages}{832--844}.
\newblock
\showISSN{0167-6393}
\urldef\tempurl%
\url{https://doi.org/10.1016/j.specom.2009.04.005}
\showDOI{\tempurl}
\newblock
\shownote{Spoken Language Technology for Education.}


\bibitem[\protect\citeauthoryear{Evanini, Hauck, and Hakuta}{Evanini
  et~al\mbox{.}}{2017}]%
        {evanini2017approaches}
\bibfield{author}{\bibinfo{person}{Keelan Evanini},
  \bibinfo{person}{Maurice~Cogan Hauck}, {and} \bibinfo{person}{Kenji Hakuta}.}
  \bibinfo{year}{2017}\natexlab{}.
\newblock \showarticletitle{Approaches to automated scoring of speaking for
  K--12 English language proficiency assessments}.
\newblock \bibinfo{journal}{\emph{ETS Research Report Series}}
  \bibinfo{volume}{2017}, \bibinfo{number}{1} (\bibinfo{year}{2017}),
  \bibinfo{pages}{1--11}.
\newblock


\bibitem[\protect\citeauthoryear{Flek}{Flek}{2020}]%
        {flek2020returning}
\bibfield{author}{\bibinfo{person}{Lucie Flek}.}
  \bibinfo{year}{2020}\natexlab{}.
\newblock \showarticletitle{Returning the N to NLP: Towards contextually
  personalized classification models}. In \bibinfo{booktitle}{\emph{Proceedings
  of the 58th Annual Meeting of the Association for Computational
  Linguistics}}. \bibinfo{pages}{7828--7838}.
\newblock


\bibitem[\protect\citeauthoryear{Gemmeke, Ellis, Freedman, Jansen, Lawrence,
  Moore, Plakal, and Ritter}{Gemmeke et~al\mbox{.}}{2017}]%
        {gemmeke2017audio}
\bibfield{author}{\bibinfo{person}{Jort~F Gemmeke}, \bibinfo{person}{Daniel~PW
  Ellis}, \bibinfo{person}{Dylan Freedman}, \bibinfo{person}{Aren Jansen},
  \bibinfo{person}{Wade Lawrence}, \bibinfo{person}{R~Channing Moore},
  \bibinfo{person}{Manoj Plakal}, {and} \bibinfo{person}{Marvin Ritter}.}
  \bibinfo{year}{2017}\natexlab{}.
\newblock \showarticletitle{Audio set: An ontology and human-labeled dataset
  for audio events}. In \bibinfo{booktitle}{\emph{2017 IEEE International
  Conference on Acoustics, Speech and Signal Processing (ICASSP)}}. IEEE,
  \bibinfo{pages}{776--780}.
\newblock


\bibitem[\protect\citeauthoryear{Grover, Kumar, Sarin, Vafaee, Hama, and
  Shah}{Grover et~al\mbox{.}}{2020}]%
        {grover2020multi}
\bibfield{author}{\bibinfo{person}{Manraj~Singh Grover}, \bibinfo{person}{Yaman
  Kumar}, \bibinfo{person}{Sumit Sarin}, \bibinfo{person}{Payman Vafaee},
  \bibinfo{person}{Mika Hama}, {and} \bibinfo{person}{Rajiv~Ratn Shah}.}
  \bibinfo{year}{2020}\natexlab{}.
\newblock \showarticletitle{Multi-modal automated speech scoring using
  attention fusion}.
\newblock \bibinfo{journal}{\emph{arXiv preprint arXiv:2005.08182}}
  (\bibinfo{year}{2020}).
\newblock


\bibitem[\protect\citeauthoryear{Higton, Leonardi, Choudhoury, Richards, Owen,
  and Sofroniou}{Higton et~al\mbox{.}}{2017}]%
        {higton2017teacher}
\bibfield{author}{\bibinfo{person}{John Higton}, \bibinfo{person}{Sarah
  Leonardi}, \bibinfo{person}{Arifa Choudhoury}, \bibinfo{person}{Neil
  Richards}, \bibinfo{person}{David Owen}, {and} \bibinfo{person}{Nicholas
  Sofroniou}.} \bibinfo{year}{2017}\natexlab{}.
\newblock \showarticletitle{Teacher workload survey 2016}.
\newblock  (\bibinfo{year}{2017}).
\newblock


\bibitem[\protect\citeauthoryear{Honnibal and Montani}{Honnibal and
  Montani}{2017}]%
        {spacy2}
\bibfield{author}{\bibinfo{person}{Matthew Honnibal} {and}
  \bibinfo{person}{Ines Montani}.} \bibinfo{year}{2017}\natexlab{}.
\newblock \bibinfo{title}{{spaCy 2}: Natural language understanding with
  {B}loom embeddings, convolutional neural networks and incremental parsing}.
  (\bibinfo{year}{2017}).
\newblock
\newblock
\shownote{To appear.}


\bibitem[\protect\citeauthoryear{Hsu, Zhang, Weiss, Zen, Wu, Wang, Cao, Jia,
  Chen, Shen, et~al\mbox{.}}{Hsu et~al\mbox{.}}{2018}]%
        {hsu2018hierarchical}
\bibfield{author}{\bibinfo{person}{Wei-Ning Hsu}, \bibinfo{person}{Yu Zhang},
  \bibinfo{person}{Ron~J Weiss}, \bibinfo{person}{Heiga Zen},
  \bibinfo{person}{Yonghui Wu}, \bibinfo{person}{Yuxuan Wang},
  \bibinfo{person}{Yuan Cao}, \bibinfo{person}{Ye Jia},
  \bibinfo{person}{Zhifeng Chen}, \bibinfo{person}{Jonathan Shen},
  {et~al\mbox{.}}} \bibinfo{year}{2018}\natexlab{}.
\newblock \showarticletitle{Hierarchical generative modeling for controllable
  speech synthesis}.
\newblock \bibinfo{journal}{\emph{arXiv preprint arXiv:1810.07217}}
  (\bibinfo{year}{2018}).
\newblock


\bibitem[\protect\citeauthoryear{Institute}{Institute}{2020}]%
        {OhioNumbers}
\bibfield{author}{\bibinfo{person}{Thomas B.~Fordham Institute}.}
  \bibinfo{year}{2020}\natexlab{}.
\newblock \bibinfo{title}{Ohio Public School Students}.
\newblock \bibinfo{howpublished}{\url{https://www.ohiobythenumbers.com/}}.
\newblock


\bibitem[\protect\citeauthoryear{Ji, Pan, Long, Li, Jiang, and Huang}{Ji
  et~al\mbox{.}}{2019}]%
        {ji2019learning}
\bibfield{author}{\bibinfo{person}{Shaoxiong Ji}, \bibinfo{person}{Shirui Pan},
  \bibinfo{person}{Guodong Long}, \bibinfo{person}{Xue Li},
  \bibinfo{person}{Jing Jiang}, {and} \bibinfo{person}{Zi Huang}.}
  \bibinfo{year}{2019}\natexlab{}.
\newblock \showarticletitle{Learning private neural language modeling with
  attentive aggregation}. In \bibinfo{booktitle}{\emph{2019 International Joint
  Conference on Neural Networks (IJCNN)}}. IEEE, \bibinfo{pages}{1--8}.
\newblock


\bibitem[\protect\citeauthoryear{King and Cook}{King and Cook}{2020}]%
        {king2020authorship}
\bibfield{author}{\bibinfo{person}{Milton King} {and} \bibinfo{person}{Paul
  Cook}.} \bibinfo{year}{2020}\natexlab{}.
\newblock \showarticletitle{Authorship Verification with Personalized Language
  Models}. In \bibinfo{booktitle}{\emph{International Conference on Text,
  Speech, and Dialogue}}. Springer, \bibinfo{pages}{248--256}.
\newblock


\bibitem[\protect\citeauthoryear{Kingma and Ba}{Kingma and Ba}{2014}]%
        {adam}
\bibfield{author}{\bibinfo{person}{Diederik~P. Kingma} {and}
  \bibinfo{person}{Jimmy Ba}.} \bibinfo{year}{2014}\natexlab{}.
\newblock \bibinfo{title}{Adam: A Method for Stochastic Optimization}.
\newblock
\newblock
\urldef\tempurl%
\url{http://arxiv.org/abs/1412.6980}
\showURL{%
\tempurl}
\newblock
\shownote{cite arxiv:1412.6980Comment: Published as a conference paper at the
  3rd International Conference for Learning Representations, San Diego, 2015.}


\bibitem[\protect\citeauthoryear{Klus{\'a}cek, Navratil, Reynolds, and
  Campbell}{Klus{\'a}cek et~al\mbox{.}}{2003}]%
        {klusacek2003conditional}
\bibfield{author}{\bibinfo{person}{David Klus{\'a}cek}, \bibinfo{person}{Jiri
  Navratil}, \bibinfo{person}{Douglas~A Reynolds}, {and}
  \bibinfo{person}{Joseph~P Campbell}.} \bibinfo{year}{2003}\natexlab{}.
\newblock \showarticletitle{Conditional pronunciation modeling in speaker
  detection}. In \bibinfo{booktitle}{\emph{2003 IEEE International Conference
  on Acoustics, Speech, and Signal Processing, 2003.
  Proceedings.(ICASSP'03).}}, Vol.~\bibinfo{volume}{4}. IEEE,
  \bibinfo{pages}{IV--804}.
\newblock


\bibitem[\protect\citeauthoryear{Kokhlikyan, Miglani, Martin, Wang, Alsallakh,
  Reynolds, Melnikov, Kliushkina, Araya, Yan, and
  Reblitz-Richardson}{Kokhlikyan et~al\mbox{.}}{2020}]%
        {captumai}
\bibfield{author}{\bibinfo{person}{Narine Kokhlikyan}, \bibinfo{person}{Vivek
  Miglani}, \bibinfo{person}{Miguel Martin}, \bibinfo{person}{Edward Wang},
  \bibinfo{person}{Bilal Alsallakh}, \bibinfo{person}{Jonathan Reynolds},
  \bibinfo{person}{Alexander Melnikov}, \bibinfo{person}{Natalia Kliushkina},
  \bibinfo{person}{Carlos Araya}, \bibinfo{person}{Siqi Yan}, {and}
  \bibinfo{person}{Orion Reblitz-Richardson}.} \bibinfo{year}{2020}\natexlab{}.
\newblock \bibinfo{title}{Captum: A unified and generic model interpretability
  library for PyTorch}.
\newblock
\newblock
\showeprint[arxiv]{2009.07896}~[cs.LG]


\bibitem[\protect\citeauthoryear{Kumar, Aggarwal, Mahata, Shah, Kumaraguru, and
  Zimmermann}{Kumar et~al\mbox{.}}{2019}]%
        {kumar2019get}
\bibfield{author}{\bibinfo{person}{Yaman Kumar}, \bibinfo{person}{Swati
  Aggarwal}, \bibinfo{person}{Debanjan Mahata}, \bibinfo{person}{Rajiv~Ratn
  Shah}, \bibinfo{person}{Ponnurangam Kumaraguru}, {and} \bibinfo{person}{Roger
  Zimmermann}.} \bibinfo{year}{2019}\natexlab{}.
\newblock \showarticletitle{Get IT Scored Using AutoSAS—An Automated System
  for Scoring Short Answers}. In \bibinfo{booktitle}{\emph{Proceedings of the
  AAAI Conference on Artificial Intelligence}}, Vol.~\bibinfo{volume}{33}.
  \bibinfo{pages}{9662--9669}.
\newblock


\bibitem[\protect\citeauthoryear{Kumar, Bhatia, Kabra, Li, Jin, and Shah}{Kumar
  et~al\mbox{.}}{2020}]%
        {kumar2020calling}
\bibfield{author}{\bibinfo{person}{Yaman Kumar}, \bibinfo{person}{Mehar
  Bhatia}, \bibinfo{person}{Anubha Kabra}, \bibinfo{person}{Jessy~Junyi Li},
  \bibinfo{person}{Di Jin}, {and} \bibinfo{person}{Rajiv~Ratn Shah}.}
  \bibinfo{year}{2020}\natexlab{}.
\newblock \showarticletitle{Calling out bluff: Attacking the robustness of
  automatic scoring systems with simple adversarial testing}.
\newblock \bibinfo{journal}{\emph{arXiv preprint arXiv:2007.06796}}
  (\bibinfo{year}{2020}).
\newblock


\bibitem[\protect\citeauthoryear{LaFlair and Settles}{LaFlair and
  Settles}{2019}]%
        {laflair2019duolingo}
\bibfield{author}{\bibinfo{person}{Geoffrey~T LaFlair} {and}
  \bibinfo{person}{Burr Settles}.} \bibinfo{year}{2019}\natexlab{}.
\newblock \showarticletitle{Duolingo English test: Technical manual}.
\newblock \bibinfo{journal}{\emph{Retrieved April}}  \bibinfo{volume}{28}
  (\bibinfo{year}{2019}), \bibinfo{pages}{2020}.
\newblock


\bibitem[\protect\citeauthoryear{Le}{Le}{2020}]%
        {ibisWorldEdTesting}
\bibfield{author}{\bibinfo{person}{Thi Le}.} \bibinfo{year}{2020}\natexlab{}.
\newblock \bibinfo{title}{Testing \& Educational Support in the US}.
\newblock
  \bibinfo{howpublished}{\url{https://my.ibisworld.com/us/en/industry/61171/key-statistics
  }}.
\newblock


\bibitem[\protect\citeauthoryear{Loukina and Cahill}{Loukina and
  Cahill}{2016}]%
        {loukina2016automated}
\bibfield{author}{\bibinfo{person}{Anastassia Loukina} {and}
  \bibinfo{person}{Aoife Cahill}.} \bibinfo{year}{2016}\natexlab{}.
\newblock \showarticletitle{Automated scoring across different modalities}. In
  \bibinfo{booktitle}{\emph{Proceedings of the 11th workshop on innovative use
  of NLP for building educational applications}}. \bibinfo{pages}{130--135}.
\newblock


\bibitem[\protect\citeauthoryear{Madnani and Cahill}{Madnani and
  Cahill}{2018}]%
        {madnani2018automated}
\bibfield{author}{\bibinfo{person}{Nitin Madnani} {and} \bibinfo{person}{Aoife
  Cahill}.} \bibinfo{year}{2018}\natexlab{}.
\newblock \showarticletitle{Automated scoring: Beyond natural language
  processing}. In \bibinfo{booktitle}{\emph{Proceedings of the 27th
  International Conference on Computational Linguistics}}.
  \bibinfo{pages}{1099--1109}.
\newblock


\bibitem[\protect\citeauthoryear{Malone}{Malone}{2000}]%
        {malone2000simulated}
\bibfield{author}{\bibinfo{person}{Margaret Malone}.}
  \bibinfo{year}{2000}\natexlab{}.
\newblock \showarticletitle{Simulated Oral Proficiency Interviews: Recent
  Developments. ERIC Digest.}
\newblock  (\bibinfo{year}{2000}).
\newblock


\bibitem[\protect\citeauthoryear{McCarthy, Puzon, and Pino}{McCarthy
  et~al\mbox{.}}{2020}]%
        {mccarthy2020skinaugment}
\bibfield{author}{\bibinfo{person}{Arya~D McCarthy}, \bibinfo{person}{Liezl
  Puzon}, {and} \bibinfo{person}{Juan Pino}.} \bibinfo{year}{2020}\natexlab{}.
\newblock \showarticletitle{SkinAugment: auto-encoding speaker conversions for
  automatic speech translation}. In \bibinfo{booktitle}{\emph{ICASSP 2020-2020
  IEEE International Conference on Acoustics, Speech and Signal Processing
  (ICASSP)}}. IEEE, \bibinfo{pages}{7924--7928}.
\newblock


\bibitem[\protect\citeauthoryear{Mudrakarta, Taly, Sundararajan, and
  Dhamdhere}{Mudrakarta et~al\mbox{.}}{2018}]%
        {mudrakarta2018did}
\bibfield{author}{\bibinfo{person}{Pramod~Kaushik Mudrakarta},
  \bibinfo{person}{Ankur Taly}, \bibinfo{person}{Mukund Sundararajan}, {and}
  \bibinfo{person}{Kedar Dhamdhere}.} \bibinfo{year}{2018}\natexlab{}.
\newblock \showarticletitle{Did the model understand the question?}
\newblock \bibinfo{journal}{\emph{arXiv preprint arXiv:1805.05492}}
  (\bibinfo{year}{2018}).
\newblock


\bibitem[\protect\citeauthoryear{O'Donnell}{O'Donnell}{2020}]%
        {ohioAES2}
\bibfield{author}{\bibinfo{person}{Patrick O'Donnell}.}
  \bibinfo{year}{2020}\natexlab{}.
\newblock \bibinfo{title}{Computers are now grading essays on Ohio's state
  tests}.
\newblock
  \bibinfo{howpublished}{\url{https://www.cleveland.com/metro/2018/03/computers_are_now_grading_essays_on_ohios_state_tests_your_ch.html}}.
\newblock


\bibitem[\protect\citeauthoryear{OECD}{OECD}{2014}]%
        {oecd2014indicator}
\bibfield{author}{\bibinfo{person}{OECD}.} \bibinfo{year}{2014}\natexlab{}.
\newblock \bibinfo{title}{Indicator D4: How much time do teachers spend
  teaching?}
\newblock
\newblock


\bibitem[\protect\citeauthoryear{Oraby, Reed, Tandon, Sharath, Lukin, and
  Walker}{Oraby et~al\mbox{.}}{2018}]%
        {oraby2018controlling}
\bibfield{author}{\bibinfo{person}{Shereen Oraby}, \bibinfo{person}{Lena Reed},
  \bibinfo{person}{Shubhangi Tandon}, \bibinfo{person}{TS Sharath},
  \bibinfo{person}{Stephanie Lukin}, {and} \bibinfo{person}{Marilyn Walker}.}
  \bibinfo{year}{2018}\natexlab{}.
\newblock \showarticletitle{Controlling personality-based stylistic variation
  with neural natural language generators}.
\newblock \bibinfo{journal}{\emph{arXiv preprint arXiv:1805.08352}}
  (\bibinfo{year}{2018}).
\newblock


\bibitem[\protect\citeauthoryear{Page}{Page}{1967}]%
        {page1967statistical}
\bibfield{author}{\bibinfo{person}{Ellis~B Page}.}
  \bibinfo{year}{1967}\natexlab{}.
\newblock \showarticletitle{Statistical and linguistic strategies in the
  computer grading of essays}. In \bibinfo{booktitle}{\emph{COLING 1967 Volume
  1: Conference Internationale Sur Le Traitement Automatique Des Langues}}.
\newblock


\bibitem[\protect\citeauthoryear{Panayotov, Chen, Povey, and
  Khudanpur}{Panayotov et~al\mbox{.}}{2015}]%
        {librispeech}
\bibfield{author}{\bibinfo{person}{Vassil Panayotov}, \bibinfo{person}{Guoguo
  Chen}, \bibinfo{person}{Daniel Povey}, {and} \bibinfo{person}{Sanjeev
  Khudanpur}.} \bibinfo{year}{2015}\natexlab{}.
\newblock \showarticletitle{Librispeech: An ASR corpus based on public domain
  audio books}. \bibinfo{pages}{5206--5210}.
\newblock
\urldef\tempurl%
\url{https://doi.org/10.1109/ICASSP.2015.7178964}
\showDOI{\tempurl}


\bibitem[\protect\citeauthoryear{Parekh, Singla, Chen, Li, and Shah}{Parekh
  et~al\mbox{.}}{2020}]%
        {parekh2020my}
\bibfield{author}{\bibinfo{person}{Swapnil Parekh},
  \bibinfo{person}{Yaman~Kumar Singla}, \bibinfo{person}{Changyou Chen},
  \bibinfo{person}{Junyi~Jessy Li}, {and} \bibinfo{person}{Rajiv~Ratn Shah}.}
  \bibinfo{year}{2020}\natexlab{}.
\newblock \showarticletitle{My Teacher Thinks The World Is Flat! Interpreting
  Automatic Essay Scoring Mechanism}.
\newblock \bibinfo{journal}{\emph{arXiv preprint arXiv:2012.13872}}
  (\bibinfo{year}{2020}).
\newblock


\bibitem[\protect\citeauthoryear{Paszke, Gross, Massa, Lerer, Bradbury, Chanan,
  Killeen, Lin, Gimelshein, Antiga, Desmaison, Kopf, Yang, DeVito, Raison,
  Tejani, Chilamkurthy, Steiner, Fang, Bai, and Chintala}{Paszke
  et~al\mbox{.}}{2019}]%
        {pytorch}
\bibfield{author}{\bibinfo{person}{Adam Paszke}, \bibinfo{person}{Sam Gross},
  \bibinfo{person}{Francisco Massa}, \bibinfo{person}{Adam Lerer},
  \bibinfo{person}{James Bradbury}, \bibinfo{person}{Gregory Chanan},
  \bibinfo{person}{Trevor Killeen}, \bibinfo{person}{Zeming Lin},
  \bibinfo{person}{Natalia Gimelshein}, \bibinfo{person}{Luca Antiga},
  \bibinfo{person}{Alban Desmaison}, \bibinfo{person}{Andreas Kopf},
  \bibinfo{person}{Edward Yang}, \bibinfo{person}{Zachary DeVito},
  \bibinfo{person}{Martin Raison}, \bibinfo{person}{Alykhan Tejani},
  \bibinfo{person}{Sasank Chilamkurthy}, \bibinfo{person}{Benoit Steiner},
  \bibinfo{person}{Lu Fang}, \bibinfo{person}{Junjie Bai}, {and}
  \bibinfo{person}{Soumith Chintala}.} \bibinfo{year}{2019}\natexlab{}.
\newblock \showarticletitle{PyTorch: An Imperative Style, High-Performance Deep
  Learning Library}.
\newblock In \bibinfo{booktitle}{\emph{Advances in Neural Information
  Processing Systems 32}}, \bibfield{editor}{\bibinfo{person}{H.~Wallach},
  \bibinfo{person}{H.~Larochelle}, \bibinfo{person}{A.~Beygelzimer},
  \bibinfo{person}{F.~d\textquotesingle Alch\'{e}-Buc},
  \bibinfo{person}{E.~Fox}, {and} \bibinfo{person}{R.~Garnett}} (Eds.).
  \bibinfo{publisher}{Curran Associates, Inc.}, \bibinfo{pages}{8024--8035}.
\newblock
\urldef\tempurl%
\url{http://papers.neurips.cc/paper/9015-pytorch-an-imperative-style-high-performance-deep-learning-library.pdf}
\showURL{%
\tempurl}


\bibitem[\protect\citeauthoryear{Patil, Singla, Shah, Hama, and
  Zimmermann}{Patil et~al\mbox{.}}{2020}]%
        {patil2020towards}
\bibfield{author}{\bibinfo{person}{Rajaswa Patil}, \bibinfo{person}{Yaman~Kumar
  Singla}, \bibinfo{person}{Rajiv~Ratn Shah}, \bibinfo{person}{Mika Hama},
  {and} \bibinfo{person}{Roger Zimmermann}.} \bibinfo{year}{2020}\natexlab{}.
\newblock \showarticletitle{Towards Modelling Coherence in Spoken Discourse}.
\newblock \bibinfo{journal}{\emph{arXiv preprint arXiv:2101.00056}}
  (\bibinfo{year}{2020}).
\newblock


\bibitem[\protect\citeauthoryear{Pennington, Socher, and Manning}{Pennington
  et~al\mbox{.}}{2014}]%
        {glove}
\bibfield{author}{\bibinfo{person}{Jeffrey Pennington},
  \bibinfo{person}{Richard Socher}, {and} \bibinfo{person}{Christopher~D.
  Manning}.} \bibinfo{year}{2014}\natexlab{}.
\newblock \showarticletitle{Glove: Global vectors for word representation}. In
  \bibinfo{booktitle}{\emph{In EMNLP}}.
\newblock


\bibitem[\protect\citeauthoryear{Povey, Ghoshal, Boulianne, Burget, Glembek,
  Goel, Hannemann, Motlicek, Qian, Schwarz, Silovsky, Stemmer, and
  Vesely}{Povey et~al\mbox{.}}{2011}]%
        {Povey_ASRU2011Kaldi}
\bibfield{author}{\bibinfo{person}{Daniel Povey}, \bibinfo{person}{Arnab
  Ghoshal}, \bibinfo{person}{Gilles Boulianne}, \bibinfo{person}{Lukas Burget},
  \bibinfo{person}{Ondrej Glembek}, \bibinfo{person}{Nagendra Goel},
  \bibinfo{person}{Mirko Hannemann}, \bibinfo{person}{Petr Motlicek},
  \bibinfo{person}{Yanmin Qian}, \bibinfo{person}{Petr Schwarz},
  \bibinfo{person}{Jan Silovsky}, \bibinfo{person}{Georg Stemmer}, {and}
  \bibinfo{person}{Karel Vesely}.} \bibinfo{year}{2011}\natexlab{}.
\newblock \showarticletitle{The Kaldi Speech Recognition Toolkit}. In
  \bibinfo{booktitle}{\emph{IEEE 2011 Workshop on Automatic Speech Recognition
  and Understanding}} (Hilton Waikoloa Village, Big Island, Hawaii, US).
  \bibinfo{publisher}{IEEE Signal Processing Society}.
\newblock
\newblock
\shownote{IEEE Catalog No.: CFP11SRW-USB.}


\bibitem[\protect\citeauthoryear{Qian, Lange, Evanini, Pugh, Ubale, Mulholland,
  and Wang}{Qian et~al\mbox{.}}{2019}]%
        {qian2019neural}
\bibfield{author}{\bibinfo{person}{Yao Qian}, \bibinfo{person}{Patrick Lange},
  \bibinfo{person}{Keelan Evanini}, \bibinfo{person}{Robert Pugh},
  \bibinfo{person}{Rutuja Ubale}, \bibinfo{person}{Matthew Mulholland}, {and}
  \bibinfo{person}{Xinhao Wang}.} \bibinfo{year}{2019}\natexlab{}.
\newblock \showarticletitle{Neural approaches to automated speech scoring of
  monologue and dialogue responses}. In \bibinfo{booktitle}{\emph{ICASSP
  2019-2019 IEEE International Conference on Acoustics, Speech and Signal
  Processing (ICASSP)}}. IEEE, \bibinfo{pages}{8112--8116}.
\newblock


\bibitem[\protect\citeauthoryear{Qian, Ubale, Mulholland, Evanini, and
  Wang}{Qian et~al\mbox{.}}{2018}]%
        {qian2018prompt}
\bibfield{author}{\bibinfo{person}{Yao Qian}, \bibinfo{person}{Rutuja Ubale},
  \bibinfo{person}{Matthew Mulholland}, \bibinfo{person}{Keelan Evanini}, {and}
  \bibinfo{person}{Xinhao Wang}.} \bibinfo{year}{2018}\natexlab{}.
\newblock \showarticletitle{A prompt-aware neural network approach to
  content-based scoring of non-native spontaneous speech}. In
  \bibinfo{booktitle}{\emph{2018 IEEE Spoken Language Technology Workshop
  (SLT)}}. IEEE, \bibinfo{pages}{979--986}.
\newblock


\bibitem[\protect\citeauthoryear{Rouvier, Bousquet, and Favre}{Rouvier
  et~al\mbox{.}}{2015}]%
        {rouvier2015speaker}
\bibfield{author}{\bibinfo{person}{Mickael Rouvier},
  \bibinfo{person}{Pierre-Michel Bousquet}, {and} \bibinfo{person}{Benoit
  Favre}.} \bibinfo{year}{2015}\natexlab{}.
\newblock \showarticletitle{Speaker diarization through speaker embeddings}. In
  \bibinfo{booktitle}{\emph{2015 23rd European Signal Processing Conference
  (EUSIPCO)}}. IEEE, \bibinfo{pages}{2082--2086}.
\newblock


\bibitem[\protect\citeauthoryear{Saeki, Matsuyama, Kobashikawa, Ogawa, and
  Kobayashi}{Saeki et~al\mbox{.}}{2021}]%
        {saeki2021analysis}
\bibfield{author}{\bibinfo{person}{Mao Saeki}, \bibinfo{person}{Yoichi
  Matsuyama}, \bibinfo{person}{Satoshi Kobashikawa}, \bibinfo{person}{Tetsuji
  Ogawa}, {and} \bibinfo{person}{Tetsunori Kobayashi}.}
  \bibinfo{year}{2021}\natexlab{}.
\newblock \showarticletitle{Analysis of Multimodal Features for Speaking
  Proficiency Scoring in an Interview Dialogue}. In
  \bibinfo{booktitle}{\emph{2021 IEEE Spoken Language Technology Workshop
  (SLT)}}. IEEE, \bibinfo{pages}{629--635}.
\newblock


\bibitem[\protect\citeauthoryear{Service}{Service}{2020}]%
        {causeIQETSWorth}
\bibfield{author}{\bibinfo{person}{Education~Testing Service}.}
  \bibinfo{year}{2020}\natexlab{}.
\newblock \bibinfo{title}{Education Testing Service EIN 21-0634479}.
\newblock
  \bibinfo{howpublished}{\url{https://www.causeiq.com/organizations/educational-testing-service,210634479/
  }}.
\newblock


\bibitem[\protect\citeauthoryear{Shah, Singla, Chen, and Shah}{Shah
  et~al\mbox{.}}{2021}]%
        {shah2021all}
\bibfield{author}{\bibinfo{person}{Jui Shah}, \bibinfo{person}{Yaman~Kumar
  Singla}, \bibinfo{person}{Changyou Chen}, {and} \bibinfo{person}{Rajiv~Ratn
  Shah}.} \bibinfo{year}{2021}\natexlab{}.
\newblock \showarticletitle{What all do audio transformer models hear? Probing
  Acoustic Representations for Language Delivery and its Structure}.
\newblock \bibinfo{journal}{\emph{arXiv preprint arXiv:2101.00387}}
  (\bibinfo{year}{2021}).
\newblock


\bibitem[\protect\citeauthoryear{Shrikumar, Greenside, Shcherbina, and
  Kundaje}{Shrikumar et~al\mbox{.}}{2016}]%
        {shrikumar2016not}
\bibfield{author}{\bibinfo{person}{Avanti Shrikumar}, \bibinfo{person}{Peyton
  Greenside}, \bibinfo{person}{Anna Shcherbina}, {and} \bibinfo{person}{Anshul
  Kundaje}.} \bibinfo{year}{2016}\natexlab{}.
\newblock \showarticletitle{Not just a black box: Learning important features
  through propagating activation differences}.
\newblock \bibinfo{journal}{\emph{arXiv preprint arXiv:1605.01713}}
  (\bibinfo{year}{2016}).
\newblock


\bibitem[\protect\citeauthoryear{SLTI}{SLTI}{2020}]%
        {sopi-slti}
\bibfield{author}{\bibinfo{person}{SLTI}.} \bibinfo{year}{2020}\natexlab{}.
\newblock \bibinfo{title}{Simulated Oral Proficiency Interview (SOPI) by SLTI}.
\newblock
  \bibinfo{howpublished}{\url{https://secondlanguagetesting.com/products-%26-services#5eb17e51-2737-458f-96a1-9101d1e453e4
  }}.
\newblock


\bibitem[\protect\citeauthoryear{Smith and Stansfield}{Smith and
  Stansfield}{2017}]%
        {smith2017testing}
\bibfield{author}{\bibinfo{person}{Megan Smith} {and}
  \bibinfo{person}{Charles~W Stansfield}.} \bibinfo{year}{2017}\natexlab{}.
\newblock \showarticletitle{Testing aptitude for second language learning}.
\newblock \bibinfo{journal}{\emph{Language Testing and Assessment}}
  (\bibinfo{year}{2017}), \bibinfo{pages}{1--14}.
\newblock


\bibitem[\protect\citeauthoryear{Stansfield and Winke}{Stansfield and
  Winke}{2008}]%
        {stansfield2008testing}
\bibfield{author}{\bibinfo{person}{Charles Stansfield} {and}
  \bibinfo{person}{Paula Winke}.} \bibinfo{year}{2008}\natexlab{}.
\newblock \showarticletitle{Testing aptitude for second language learning}.
\newblock \bibinfo{journal}{\emph{Encyclopaedia of language and education, 2nd
  Edition: Language Testing and assessment}}  \bibinfo{volume}{7}
  (\bibinfo{year}{2008}), \bibinfo{pages}{81--94}.
\newblock


\bibitem[\protect\citeauthoryear{Stansfield and Kenyon}{Stansfield and
  Kenyon}{1992a}]%
        {stansfield1992development}
\bibfield{author}{\bibinfo{person}{Charles~W Stansfield} {and}
  \bibinfo{person}{Dorry~Mann Kenyon}.} \bibinfo{year}{1992}\natexlab{a}.
\newblock \showarticletitle{The development and validation of a simulated oral
  proficiency interview}.
\newblock \bibinfo{journal}{\emph{The Modern Language Journal}}
  \bibinfo{volume}{76}, \bibinfo{number}{2} (\bibinfo{year}{1992}),
  \bibinfo{pages}{129--141}.
\newblock


\bibitem[\protect\citeauthoryear{Stansfield and Kenyon}{Stansfield and
  Kenyon}{1992b}]%
        {stansfield1992research}
\bibfield{author}{\bibinfo{person}{Charles~W Stansfield} {and}
  \bibinfo{person}{Dorry~Mann Kenyon}.} \bibinfo{year}{1992}\natexlab{b}.
\newblock \showarticletitle{Research on the comparability of the oral
  proficiency interview and the simulated oral proficiency interview}.
\newblock \bibinfo{journal}{\emph{System}} \bibinfo{volume}{20},
  \bibinfo{number}{3} (\bibinfo{year}{1992}), \bibinfo{pages}{347--364}.
\newblock


\bibitem[\protect\citeauthoryear{Stansfield and Kenyon}{Stansfield and
  Kenyon}{1996}]%
        {stansfield1996test}
\bibfield{author}{\bibinfo{person}{Charles~W Stansfield} {and}
  \bibinfo{person}{Dorry~Mann Kenyon}.} \bibinfo{year}{1996}\natexlab{}.
\newblock \bibinfo{booktitle}{\emph{Test Development Handbook: Simulated Oral
  Proficiency Interview,(SOPI)}}.
\newblock \bibinfo{publisher}{Center for Applied Linguistics}.
\newblock


\bibitem[\protect\citeauthoryear{Strauss}{Strauss}{2020}]%
        {educationTestingSalaries}
\bibfield{author}{\bibinfo{person}{Valerie Strauss}.}
  \bibinfo{year}{2020}\natexlab{}.
\newblock \bibinfo{title}{How much do big education nonprofits pay their
  bosses? Quite a bit, it turns out.}
\newblock
  \bibinfo{howpublished}{\url{https://www.washingtonpost.com/news/answer-sheet/wp/2015/09/30/how-much-do-big-education-nonprofits-pay-their-bosses-quite-a-bit-it-turns-out/}}.
\newblock


\bibitem[\protect\citeauthoryear{Sundararajan, Taly, and Yan}{Sundararajan
  et~al\mbox{.}}{2017}]%
        {sundararajan2017axiomatic}
\bibfield{author}{\bibinfo{person}{Mukund Sundararajan}, \bibinfo{person}{Ankur
  Taly}, {and} \bibinfo{person}{Qiqi Yan}.} \bibinfo{year}{2017}\natexlab{}.
\newblock \showarticletitle{Axiomatic attribution for deep networks}.
\newblock \bibinfo{journal}{\emph{arXiv preprint arXiv:1703.01365}}
  (\bibinfo{year}{2017}).
\newblock


\bibitem[\protect\citeauthoryear{Taghipour and Ng}{Taghipour and Ng}{2016}]%
        {taghipour-ng-2016-neural}
\bibfield{author}{\bibinfo{person}{Kaveh Taghipour} {and}
  \bibinfo{person}{Hwee~Tou Ng}.} \bibinfo{year}{2016}\natexlab{}.
\newblock \showarticletitle{A Neural Approach to Automated Essay Scoring}. In
  \bibinfo{booktitle}{\emph{Proceedings of the 2016 Conference on Empirical
  Methods in Natural Language Processing}}. \bibinfo{publisher}{Association for
  Computational Linguistics}, \bibinfo{address}{Austin, Texas},
  \bibinfo{pages}{1882--1891}.
\newblock
\urldef\tempurl%
\url{https://doi.org/10.18653/v1/D16-1193}
\showDOI{\tempurl}


\bibitem[\protect\citeauthoryear{Tay, Phan, Tuan, and Hui}{Tay
  et~al\mbox{.}}{2017}]%
        {tay}
\bibfield{author}{\bibinfo{person}{Yi Tay}, \bibinfo{person}{Minh Phan},
  \bibinfo{person}{Luu Tuan}, {and} \bibinfo{person}{Siu Hui}.}
  \bibinfo{year}{2017}\natexlab{}.
\newblock \showarticletitle{SkipFlow: Incorporating Neural Coherence Features
  for End-to-End Automatic Text Scoring}.
\newblock  (\bibinfo{date}{11} \bibinfo{year}{2017}).
\newblock


\bibitem[\protect\citeauthoryear{TechNavio}{TechNavio}{2020}]%
        {researchAndMarketsEdTesting}
\bibfield{author}{\bibinfo{person}{TechNavio}.}
  \bibinfo{year}{2020}\natexlab{}.
\newblock \bibinfo{title}{Global Higher Education Testing and Assessment Market
  2020-2024}.
\newblock
  \bibinfo{howpublished}{\url{https://www.researchandmarkets.com/reports/5136950/global-higher-education-testing-and-assessment
  }}.
\newblock


\bibitem[\protect\citeauthoryear{Thomas}{Thomas}{2016}]%
        {thomas2016future}
\bibfield{author}{\bibinfo{person}{Susan Thomas}.}
  \bibinfo{year}{2016}\natexlab{}.
\newblock \showarticletitle{Future Ready Learning: Reimagining the Role of
  Technology in Education. 2016 National Education Technology Plan.}
\newblock \bibinfo{journal}{\emph{Office of Educational Technology, US
  Department of Education}} (\bibinfo{year}{2016}).
\newblock


\bibitem[\protect\citeauthoryear{USBE}{USBE}{2020}]%
        {UtahNumbers}
\bibfield{author}{\bibinfo{person}{USBE}.} \bibinfo{year}{2020}\natexlab{}.
\newblock \bibinfo{title}{UTAH STATE BOARD OF EDUCATION 2018–19 FINGERTIP
  FACTS}.
\newblock
  \bibinfo{howpublished}{\url{https://www.ets.org/s/gre/pdf/gre_guide_table1a.pdf}}.
\newblock


\bibitem[\protect\citeauthoryear{Vaswani, Shazeer, Parmar, Uszkoreit, Jones,
  Gomez, Kaiser, and Polosukhin}{Vaswani et~al\mbox{.}}{2017}]%
        {attention}
\bibfield{author}{\bibinfo{person}{Ashish Vaswani}, \bibinfo{person}{Noam
  Shazeer}, \bibinfo{person}{Niki Parmar}, \bibinfo{person}{Jakob Uszkoreit},
  \bibinfo{person}{Llion Jones}, \bibinfo{person}{Aidan~N Gomez},
  \bibinfo{person}{\L~ukasz Kaiser}, {and} \bibinfo{person}{Illia Polosukhin}.}
  \bibinfo{year}{2017}\natexlab{}.
\newblock \showarticletitle{Attention is All you Need}. In
  \bibinfo{booktitle}{\emph{Advances in Neural Information Processing
  Systems}}, \bibfield{editor}{\bibinfo{person}{I.~Guyon},
  \bibinfo{person}{U.~V. Luxburg}, \bibinfo{person}{S.~Bengio},
  \bibinfo{person}{H.~Wallach}, \bibinfo{person}{R.~Fergus},
  \bibinfo{person}{S.~Vishwanathan}, {and} \bibinfo{person}{R.~Garnett}}
  (Eds.), Vol.~\bibinfo{volume}{30}. \bibinfo{publisher}{Curran Associates,
  Inc.}
\newblock
\urldef\tempurl%
\url{https://proceedings.neurips.cc/paper/2017/file/3f5ee243547dee91fbd053c1c4a845aa-Paper.pdf}
\showURL{%
\tempurl}


\bibitem[\protect\citeauthoryear{Wang, Evanini, Qian, and Mulholland}{Wang
  et~al\mbox{.}}{2021}]%
        {wang2021automated}
\bibfield{author}{\bibinfo{person}{Xinhao Wang}, \bibinfo{person}{Keelan
  Evanini}, \bibinfo{person}{Yao Qian}, {and} \bibinfo{person}{Matthew
  Mulholland}.} \bibinfo{year}{2021}\natexlab{}.
\newblock \showarticletitle{Automated Scoring of Spontaneous Speech from Young
  Learners of English Using Transformers}. In \bibinfo{booktitle}{\emph{2021
  IEEE Spoken Language Technology Workshop (SLT)}}. IEEE,
  \bibinfo{pages}{705--712}.
\newblock


\bibitem[\protect\citeauthoryear{Wang, Wei, Zhou, and Huang}{Wang
  et~al\mbox{.}}{2018}]%
        {wang2018automatic}
\bibfield{author}{\bibinfo{person}{Yucheng Wang}, \bibinfo{person}{Zhongyu
  Wei}, \bibinfo{person}{Yaqian Zhou}, {and} \bibinfo{person}{Xuan-Jing
  Huang}.} \bibinfo{year}{2018}\natexlab{}.
\newblock \showarticletitle{Automatic essay scoring incorporating rating schema
  via reinforcement learning}. In \bibinfo{booktitle}{\emph{Proceedings of the
  2018 conference on empirical methods in natural language processing}}.
  \bibinfo{pages}{791--797}.
\newblock


\bibitem[\protect\citeauthoryear{Witt and Young}{Witt and Young}{2000}]%
        {WITT200095}
\bibfield{author}{\bibinfo{person}{S.M Witt} {and} \bibinfo{person}{S.J
  Young}.} \bibinfo{year}{2000}\natexlab{}.
\newblock \showarticletitle{Phone-level pronunciation scoring and assessment
  for interactive language learning}.
\newblock \bibinfo{journal}{\emph{Speech Communication}} \bibinfo{volume}{30},
  \bibinfo{number}{2} (\bibinfo{year}{2000}), \bibinfo{pages}{95--108}.
\newblock
\showISSN{0167-6393}
\urldef\tempurl%
\url{https://doi.org/10.1016/S0167-6393(99)00044-8}
\showDOI{\tempurl}


\bibitem[\protect\citeauthoryear{Wolf, Debut, Sanh, Chaumond, Delangue, Moi,
  Cistac, Rault, Louf, Funtowicz, and Brew}{Wolf et~al\mbox{.}}{2019}]%
        {huggingface}
\bibfield{author}{\bibinfo{person}{Thomas Wolf}, \bibinfo{person}{Lysandre
  Debut}, \bibinfo{person}{Victor Sanh}, \bibinfo{person}{Julien Chaumond},
  \bibinfo{person}{Clement Delangue}, \bibinfo{person}{Anthony Moi},
  \bibinfo{person}{Pierric Cistac}, \bibinfo{person}{Tim Rault},
  \bibinfo{person}{R{\'{e}}mi Louf}, \bibinfo{person}{Morgan Funtowicz}, {and}
  \bibinfo{person}{Jamie Brew}.} \bibinfo{year}{2019}\natexlab{}.
\newblock \showarticletitle{HuggingFace's Transformers: State-of-the-art
  Natural Language Processing}.
\newblock \bibinfo{journal}{\emph{CoRR}}  \bibinfo{volume}{abs/1910.03771}
  (\bibinfo{year}{2019}).
\newblock
\showeprint[arxiv]{1910.03771}
\urldef\tempurl%
\url{http://arxiv.org/abs/1910.03771}
\showURL{%
\tempurl}


\bibitem[\protect\citeauthoryear{Xiong, Evanini, Zechner, and Chen}{Xiong
  et~al\mbox{.}}{2013}]%
        {xiong2013automated}
\bibfield{author}{\bibinfo{person}{Wenting Xiong}, \bibinfo{person}{Keelan
  Evanini}, \bibinfo{person}{Klaus Zechner}, {and} \bibinfo{person}{Lei Chen}.}
  \bibinfo{year}{2013}\natexlab{}.
\newblock \showarticletitle{Automated content scoring of spoken responses
  containing multiple parts with factual information}. In
  \bibinfo{booktitle}{\emph{Speech and Language Technology in Education}}.
\newblock


\bibitem[\protect\citeauthoryear{Yan, Rupp, and Foltz}{Yan
  et~al\mbox{.}}{2020}]%
        {yan2020handbook}
\bibfield{author}{\bibinfo{person}{Duanli Yan}, \bibinfo{person}{Andr{\'e}~A
  Rupp}, {and} \bibinfo{person}{Peter~W Foltz}.}
  \bibinfo{year}{2020}\natexlab{}.
\newblock \bibinfo{booktitle}{\emph{Handbook of automated scoring: Theory into
  practice}}.
\newblock \bibinfo{publisher}{CRC Press}.
\newblock


\bibitem[\protect\citeauthoryear{Yang and Eisenstein}{Yang and
  Eisenstein}{2017}]%
        {yang2017overcoming}
\bibfield{author}{\bibinfo{person}{Yi Yang} {and} \bibinfo{person}{Jacob
  Eisenstein}.} \bibinfo{year}{2017}\natexlab{}.
\newblock \showarticletitle{Overcoming language variation in sentiment analysis
  with social attention}.
\newblock \bibinfo{journal}{\emph{Transactions of the Association for
  Computational Linguistics}}  \bibinfo{volume}{5} (\bibinfo{year}{2017}),
  \bibinfo{pages}{295--307}.
\newblock


\bibitem[\protect\citeauthoryear{Zechner, Higgins, Xi, and Williamson}{Zechner
  et~al\mbox{.}}{2009}]%
        {speechrater}
\bibfield{author}{\bibinfo{person}{Klaus Zechner}, \bibinfo{person}{Derrick
  Higgins}, \bibinfo{person}{Xiaoming Xi}, {and} \bibinfo{person}{David~M.
  Williamson}.} \bibinfo{year}{2009}\natexlab{}.
\newblock \showarticletitle{Automatic Scoring of Non-Native Spontaneous Speech
  in Tests of Spoken English}.
\newblock \bibinfo{journal}{\emph{Speech Commun.}} \bibinfo{volume}{51},
  \bibinfo{number}{10} (\bibinfo{date}{Oct.} \bibinfo{year}{2009}),
  \bibinfo{pages}{883–895}.
\newblock
\showISSN{0167-6393}
\urldef\tempurl%
\url{https://doi.org/10.1016/j.specom.2009.04.009}
\showDOI{\tempurl}


\end{thebibliography}


\end{document}